\documentclass[12pt]{iopart}
\usepackage{graphics,graphicx,subfigure,adjustbox,amssymb,xcolor}

\begin{document}

\title[1D fluids with nearest-neighbour interactions]{1D fluids with repulsive nearest-neighbour interactions: Low-temperature anomalies}  

\author{Igor Trav\v{e}nec and Ladislav \v{S}amaj} 

\address{Institute of Physics, Slovak Academy of Sciences, 
D\'ubravsk\'a cesta 9, SK-84511 Bratislava, Slovakia}
\ead{Igor.Travenec@savba.sk,Ladislav.Samaj@savba.sk}
\vspace{10pt}
\begin{indented}
\item[]
\end{indented}

\begin{abstract}
A limited number of 2D and 3D materials under a constant pressure
contract in volume upon heating isobarically; this anomalous phenomenon
is known as the negative thermal expansion (NTE).
In this paper, the NTE anomaly is observed in 1D fluids
of classical particles interacting pairwisely with two competing length
scales: the hard-core diameter $a$ and the finite range $a'>a$ of
a soft repulsive potential component.
If $a'\le 2a$, the pair interactions reduce themselves to nearest
neighbours which permits a closed-form solution of thermodynamics
in the isothermal-isobaric ensemble characterized by temperature $T$
and pressure $p$.
We focus on the equation of state (EoS) which relates the average distance
between particles (reciprocal density) $l$ to $T$ and $p\ge 0$.
The EoS is expressible explicitly in terms of elementary or special
functions for specific, already known and new, cases
like the square shoulder, the linear and quadratic ramps
as well as certain types of logarithmic interaction potentials.
The emphasis is put on low-$T$ anomalies of the EoS.
Firstly, the equidistant ground state as the function of the pressure
can exhibit, at some ``compressibility'' pressures, a jump in chain spacing
from $a'$ to $a$.
Secondly, the analytical structure of the low-$T$ expansion of $l(T,p)$
depends on ranges of $p$-values.
Thirdly, the presence of the NTE anomaly depends very much
on the shape of the core-softened potential.
\end{abstract}


\vspace{2pc}


\submitto{\JPA}

\maketitle

\renewcommand{\theequation}{1.\arabic{equation}}
\setcounter{equation}{0}

\section{Introduction} \label{Sec1}
The vast majority of materials under a constant pressure expand
in volume upon heating.
However, there exist specific materials with negative thermal
expansion (NTE) which contract in volume on heating.
A well-known example of such systems is ice under the atmospheric pressure
which undergoes NTE at very low temperatures $\lesssim 63 K$
\cite{Rottger94,Evans99} as well as upon melting into liquid water
in the temperature range $0-3.98$ $^\circ C$ as a result of
opposing directional forces due to thermal kinetic expansion and hydrogen
bonding \cite{Greenwood97,Errington01}.
NTE was observed in other low-density substances with anisotropic
intermolecular potential like graphene \cite{Yoon11} and complex compounds,
namely zirconium tungstate Zr${\rm W}_2{\rm O}_8$ \cite{Mary96},
diamond and zinc-blende semiconductors \cite{Biernacki89},
${\rm Lu}_2{\rm W}_3{\rm O}_{12}$ \cite{Forster98}, etc.,
see review \cite{Lind12}.
The correlation between NTE and pressure-induced softening, observed
in molecular dynamics simulations of cubic siliceous zeolites, was suggested
to be a common feature of NTE materials \cite{Fang13}.

Rechtsmann, Stillinger and Torquato showed that NTE can occur also
in two-dimensional (2D) and three-dimensional (3D) single-component
systems with {\em isotropic} pair interactions $\phi(r)$ \cite{Rechtsman07}.
The interaction potential is supposed to have a basin of attraction
with a strongly positive third derivative $\phi'''(r)$, i.e., increasing
curvature.
This causes that the curvature to the right of a minimum is much greater
than that to the left, with a natural tendency of particles to decrease their
mutual distance under thermal fluctuations.
The necessary and sufficient conditions for NTE in these isotropic
many-particle systems were discussed in \cite{Kuzkin14}.
In 1D, it must hold that
\begin{equation} \label{Kuzkin}
\phi'''(l)>0 ,  
\end{equation}
where $l$ is the average distance between particles.
There exists another family of 2D classical many-particle models
which exhibits NTE \cite{Batten09a,Batten09b}.
Systems are composed of identical particles with isotropic soft interaction
potential which is repulsive, bounded and absolutely integrable,
with certain conditions on the Fourier component of the interaction potential.
The anomalous NTE arises there at low temperatures due to the special
topography of the energy landscape which implies ground states
ranging from fully disordered to crystalline ones.
NTE was observed also in 3D lattice models with multiple atoms per cell
and first- as well second-nearest neighbour interactions \cite{Schick16},
experiments with cubic perovskite ionic insulator ${\rm ScF}_3$
\cite{Roekeghem16}, near structural quantum phase transitions in
mercurous halides ${\rm Hg}_2{\rm X}_2$ (X=Cl, Br and I) \cite{Occhialini17},
etc., see monograph \cite{Fisher18}.
In all cases above, NTE takes place in the crystal phase of materials, below
the melting temperature.

Using computer simulations, the NTE anomaly was detected
in {\em fluid} phase of 3D spherically symmetric purely repulsive potentials
with two competing length scales: the hard-core diameter $a$ (dominant at
higher temperatures and pressures) and a soft-core range $a'>a$
(dominant at lower temperature and pressures) at which the potential is
nonanalytic
\cite{Stillinger97,Sadr98,Velasco00,Ryzhov03,Kumar05,Xu06,Yan06,Oliveira08,Fomin08,Oliveira09,Buldyrev09,Gribova09,Zhou09,Vilaseca11,Coquand20}.
The repulsive core-softened potentials used in simulations were mainly
the (square) shoulder potential introduced by Hemmer and Stell
\cite{Hemmer70,Stell72} and the (linear) ramp potential of Jagla
\cite{Jagla99}, or potentials ranging between them \cite{Oliveira08,Oliveira09}.
A controversy about the observed absence of the NTE anomaly for
the square shoulder potential \cite{Oliveira08} was later rejected
in Refs. \cite{Fomin08,Oliveira09} by arguing that waterlike anomalies
can be overlayed by crystalline phases which are more stable than liquid
in certain regions of thermodynamic parameters.
The conclusion suggested in \cite{Oliveira09} was that the two-length-scales
interaction potentials always exhibit water-like anomalies.

Exactly solvable models are very valuable in statistical mechanics 
because they can serve as tests for hypothesis, approximations and
numerical methods.
There exists a large family of 1D fluids of classical 
particles interacting pairwisely via a potential chosen so that
the interaction of a given particle with its nearest neighbours
only effectively occurs.
The exact solution of thermal equilibrium at temperature $T$
for such models involves the Equation of State (EoS), thermodynamic
potentials, the structure factor and the distribution functions.
In this paper, testing various interaction potentials we find out
that, under certain conditions, the NTE anomaly is surprisingly present
within this relatively simple family of 1D fluids.

The simplest 1D model with nearest-neighbour interactions is the Tonks gas
of hard rods with the interaction potential
\begin{equation} \label{hr}
\phi(x) = \left\{
\begin{array}{ll}
\infty & \mbox{if $\vert x\vert < a$,} \cr
0 & \mbox{if $\vert x\vert \ge a$,}  
\end{array} \right.   
\end{equation}
where $a$ is the diameter of the hard core around each particle.
Its EoS, relating the particle density with temperature $T$ and pressure
$p\ge 0$, was derived by Tonks \cite{Tonks36}. 
It is interesting that the two-particle distribution function of hard rods
was calculated much sooner by Zernike and Prins \cite{Zernike27}.

A more general model with interactions among nearest neighbours only was
solved by Takahashi \cite{Takahashi42} by using the canonical ensemble.
A simplified rederivation of the EoS by using the isothermal-isobaric ensemble,
characterized by temperature $T$ and pressure $p$,
was made by Bishop and Boonstra \cite{Bishop83}.
The many-particle distribution functions for Takahashi gas have been
computed by Salsburg, Zwanzig and Kirkwood \cite{Salsburg53}.
An alternative formula for the two-particle distribution function
in terms of thermodynamic quantities was given by Lebowitz, Percus and
Zucker \cite{Lebowitz62}.

Takahashi's model was approximative in the sense that interactions among
all pairs of particles were reduced to the nearest neighbour pairs of
particles ``by hand''.
Let us consider the following pair interaction potential
\begin{equation} \label{general}
\phi(x) = \left\{
\begin{array}{ll}
\infty & \mbox{if $\vert x\vert < a$,} \cr
\varphi(x) & \mbox{if $a\le \vert x\vert < a'$,} \cr
0 & \mbox{if $\vert x\vert \ge a'$,}  
\end{array} \right.   
\end{equation}  
where $\varphi(x)=\varphi(-x)$ and the range $a'$ of the potential
(beyond which it vanishes) is restricted to $a'\le 2a$.
This condition prevents from the interaction of a given particle with
its second nearest neighbours due to the hard cores of the first
nearest neighbours which leads to an effective reduction of the interaction
to nearest neighbours in a natural way ``without any ad-hoc assumptions''.
The inequality $a'\le 2a$ will always be considered throughout this article.
The exact solution of 1D fluids with the interaction (\ref{general})
involves many-body systems of identical particles
\cite{Salsburg53} as well as mixtures, including nonadditive hard rods
\cite{Lebowitz71,Heying04,Santos07,Ben-Naim09,Sahnoun24},
see review \cite{Percus87} and monograph \cite{Santos16}.

Surface effects of fluids with nearest-neighbour interactions
were studied in \cite{Felderhof70}.
An approximation which is exact in 1D and works well also in higher
dimensions was developed in \cite{Bakhti21}. 
Although 1D systems with finite-range interactions do not exhibit
phase transitions at nonzero temperatures \cite{Hove50}, the low-density and
$p,T\to 0$ scaling regime of a 1D fluid with attractive nearest-neighbour
interactions was shown to undertake a second-order phase transition with
long-ranged decay of correlations \cite{Jones86}.

There exist 1D fluids of pointlike particles with long-range
interactions which admit the exact solution, like the one-component plasma
(or the jellium) \cite{Baxter63}, the two-component plasma (or the
symmetric Coulomb gas) \cite{Lenard61,Edwards62} and the Kac-Backer model
\cite{Kac59,Backer61}.
Other 1D fluids with finite-range interactions going beyond the nearest
neighbours are not exactly solvable and one has to constrain oneself
to reasonable approximations \cite{Fantoni10,Fantoni17}.

In spite of a huge number of works dealing with 1D fluids with
nearest-neighbour interactions, we did not find a detailed analysis
of exact formulas in the limit of low temperatures.
This is probably caused by the general belief that the ground state
of these models is simple and nothing exceptional can happen in
their thermal equilibrium.
The present paper shows that this is not the case.

We focus on the EoS, namely in the form of the average distance between
particles (reciprocal density) $l$ as the function of $T$ and $p\ge 0$.
It admits an exact closed-form solution in terms of the Laplace transform of
the pair Boltzmann factor.
For specific models of the family (\ref{general}), this Laplace transform
is expressible explicitly in terms of elementary or special functions.
In this paper, we restrict ourselves to a few, already known or new, 
repulsive interaction potentials like the square shoulder, the linear and
quadratic ramps as well as certain types of logarithmic interaction
potentials.
For the sake of completeness, the low-$T$ and high-$T$ expansions of $l$
and the virial expansion of $p$ in integer powers of the particle density $n$
are developed for every potential.
The low-$T$ analysis of $l(T,p)$ reveals some unexpected phenomena like:

the equidistant ground state can exhibit a discontinuous jump in chain
spacing from $a$ to $a'$ at some ``compressibility'' pressure, via a degenerate
ground state when all pairs of neighbouring particles have spacing
either $a$ or $a'$ (in one to one ratio);

the analytical structure of the low-$T$ expansion of $l$ depends on
the value of the pressure;

for certain intervals of $p$-values and in short intervals of low $T$,
$l$ decreases with increasing $T$ which signals the anomalous NTE phenomenon;

the presence of the NTE anomaly depends very much on the
shape of the core-softened potential;

for certain intervals of pressure $p$, increasing $T$ from
0 the NTE anomaly may be absent in a tiny interval of $T$-values and then takes
place in an interval of strictly positive temperatures.

The paper is organized as follows.
Section \ref{Sec2} concerns the general formalism for exactly solvable
1D fluids with nearest-neighbour interactions, at arbitrary $T$ and
in the ground state at $T=0$.
The EoS of the Tonks model (\ref{hr}) is presented as the simplest example,
first few terms of the high-temperature expansion of $l$ are derived
and the compressibility factor is introduced.
The square shoulder model with the constant potential in between $a$ and $a'$
is analysed in section \ref{Sec3}.
The repulsive linear, quadratic, linear-quadratic fusion
and square-root ramp models are the subject of section \ref{Sec4}.
Section \ref{Sec5} deals with three types of interaction potentials
of logarithmic type. 
Section \ref{Sec6} is a recapitulation with concluding remarks.

\renewcommand{\theequation}{2.\arabic{equation}}
\setcounter{equation}{0}

\section{General 1D formalism} \label{Sec2}

\subsection{Exact formulas for the EoS} \label{Sec2.1}
$N$ identical particles of the 1D classical fluid move in a box of length $L$,
say with periodic boundary conditions.
The thermodynamic limit $L\to\infty$ and $N\to\infty$ while keeping
the particle number density $n=N/L$ fixed, is considered.
The pair potential $\phi(x)$ is of the form (\ref{general})
with the condition $a'\le 2 a$ which ensures that each particle interacts
only with its nearest neighbours.
For the hard-core diameter $a>0$, the condition
$\lim_{x\to 0}\phi(x)=\infty$ fixes the order of the particles on the line.

The system of particles is in thermal equilibrium at temperature $T$, 
or inverse temperature $\beta=1/(k_{\rm B}T)$ with $k_{\rm B}$
being the Boltzmann constant; we shall work in units of $k_{\rm B}=1$.
The model is exactly solvable in the isothermal-isobaric ensemble,
see monograph \cite{Santos16} for the notation.
The exact solution for the EoS can be written in terms of
the Laplace transform of the pair Boltzmann factor ${\rm e}^{-\beta\phi(x)}$,
\begin{equation}\label{Om}
\widehat{\Omega}(s) = \int_0^{\infty} {\rm d}x\,
{\rm e}^{-xs} {\rm e}^{-\beta\phi(x)} 
\end{equation}
and of its derivative
\begin{equation}
\widehat{\Omega}'(s) \equiv \frac{\partial\widehat{\Omega}(s)}{\partial s}
= - \int_0^{\infty} {\rm d}x\, x {\rm e}^{-xs} {\rm e}^{-\beta\phi(x)}
\end{equation}  
as follows
\begin{equation}\label{EoS}
n(\beta,p) = - \frac{\widehat{\Omega}(\beta p)}{\widehat{\Omega}'(\beta p)}
= \frac{\int_0^{\infty} {\rm d}x\, {\rm e}^{-x\beta p} {\rm e}^{-\beta\phi(x)}}{
\int_0^{\infty} {\rm d}x\, x {\rm e}^{-x\beta p} {\rm e}^{-\beta\phi(x)}} . 
\end{equation}  
We prefer to work with the averaged distance between particle
(reciprocal density) $l=1/n$ given by
\begin{equation}\label{recdensity}
l(\beta,p) = - \frac{\widehat{\Omega}'(\beta p)}{\widehat{\Omega}(\beta p)}
= \frac{\int_0^{\infty} {\rm d}x\, x {\rm e}^{-x\beta p} {\rm e}^{-\beta\phi(x)}}{
\int_0^{\infty} {\rm d}x\, {\rm e}^{-x\beta p} {\rm e}^{-\beta\phi(x)}} . 
\end{equation}  

\subsection{Ground state} \label{Sec2.2}
Proving the periodicity of the ground state of classical particles
interacting pairwisely is a difficult task, see Refs.
\cite{Radin87,Bris05,Blanc15} for the crystallization problem. 
A number of theorems have been proved in \cite{Ventevogel78} about
the ground state of 1D systems of classical particle interacting 
via purely repulsive potentials. 
In one of his theorems Ventevogel proved an equidistant ground state in
the case of {\em convex} repulsive potentials, for any particle density $n>0$.
Subsequently, the same result has been obtained for certain
{\em non-convex} repulsive potentials \cite{Ventevogel79a,Ventevogel79b}.

In the ground state, the pressure exerted on the $N$-particle system is given
by the change of the energy $E_0$ with respect to the length $L$: 
\begin{equation} \label{press}
p = - \frac{\partial E_0}{\partial L}
= - \frac{1}{N} \frac{\partial E_0}{\partial l} ,
\end{equation}
where the average particle distance $l=L/N$.
In the present case of convex repulsive potentials and pressures $p\ge 0$,
the spacing of the equidistant ground state $l\in [a,a']$.
If the interaction potential $\phi(x)$ is an analytic function of
$x\in (a,a')$, putting the energy of the equidistant chain of interacting
nearest neighbours $E_0 = \phi(l) N$ into (\ref{press}) results in
\begin{equation} \label{derener}
p = - \frac{\partial\phi(l)}{\partial l}.
\end{equation}
If the interaction potential $\phi(x)$ is not analytic in the interval
$x\in (a,a')$, the ground state can be determined alternatively as
the $T\to 0$ limit of the formula (\ref{recdensity}).

In connection with the ground state, we introduce the ``incompressibility''
pressure $p_i$ as the lowest pressure at which $l(T=0)=a$.

\subsection{High-temperature expansion} \label{Sec2.3}
The high-temperature expansion of $l(\beta,p)$ follows directly
from the formula (\ref{recdensity}) by expanding the exponentials
under integration in powers of $\beta$.
After simple algebra, the final result for the general model
with the interaction potential (\ref{general}) reads as
\begin{eqnarray}\label{HT}
l & = & \frac{1}{\beta p} + a + \beta \int_a^{a'} {\rm d}x\, \varphi(x)
+ \frac{\beta^2}{2} \left[ 4 a p \int_a^{a'} {\rm d}x\, \varphi(x)
\right. \nonumber \\ & & \left.  
- 4 p \int_a^{a'} {\rm d}x\, x \varphi(x) -
\int_a^{a'} {\rm d}x\, \varphi^2(x) \right] + O(\beta^3) ,  
\end{eqnarray}  
where the first term on the rhs corresponds to the ideal gas, the second
one to the excluded volume and the third term does not depend on $p$.

\subsection{Virial expansion and the compressibility factor}
A common way to treat the EoS is the virial expansion which relates
the pressure to (usually integer) powers of the particle number density:
\begin{equation} \label{Virhr}
\beta p =\sum_{k=1}^\infty B_kn^k , \qquad B_1=1 .
\end{equation}
The virial coefficients $\{ B_k \}_{k=2}^{\infty}$ depend on parameters of
the interaction potential $\varphi(x)$ and the inverse temperature $\beta$.

The deviation of thermodynamic behaviour of real gases from the ideal gas
is measured by the compressibility (or compression) factor $Z$ defined by
\begin{equation} \label{Z}
Z(\beta,n) \equiv \frac{\beta p}{n} = 1 + \sum_{k=1}^\infty B_{k+1}n^k .
\end{equation}
In the low-density limit $n\to 0$,
\begin{equation}
Z(\beta,n) \mathop{\sim}_{n\to 0} 1 .
\end{equation}
Another regime which can be treated rigorously is the close packed limit
$l\to a^+$ $(n a\to 1^-)$ when $\beta p\to \infty$.
Under appropriate shifts in $x$-variable, the exact formula (\ref{recdensity})
takes the form
\begin{equation} \label{Zl}
l = \frac{\int_0^{a'-a} {\rm d}x\, (x+a) \e^{-\beta\varphi(x+a)-\beta px} +
\e^{-(a'-a)\beta p} \int_0^{\infty} {\rm d}x\, (x+a') \e^{-\beta px}}{
\int_0^{a'-a} {\rm d}x\, \e^{-\beta\varphi(x+a)-\beta px} + \e^{-(a'-a)\beta p}
\int_0^{\infty} {\rm d}x\, \e^{-\beta px}} .
\end{equation}
Since we shall expand the rhs of this equation in inverse powers of
$1/(\beta p)$, the terms containing the exponentially small factor
$\e^{-(a'-a)\beta p}$ can be neglected in the limit $\beta p\to \infty$.
Performing in the above integrals the change of variables $y=\beta p x$,
one gets 
\begin{equation}
l = \frac{\int_0^{\beta p(a'-a)} {\rm d}y\, \left( a + \frac{y}{\beta p}\right)
\e^{-y-\beta\varphi\left( a+\frac{y}{\beta p}\right)}}{\int_0^{\beta p(a'-a)} {\rm d}y\,
\e^{-y-\beta\varphi\left( a+\frac{y}{\beta p}\right)}} .
\end{equation}
The substitution of the upper limit of the integrals
$\beta p (a'-a)$ by $\infty$ produces exponentially small terms
$\e^{-(a'-a)\beta p}$ which can be neglected one more time.
The analytic $\varphi$-function can be expanded in Taylor series around
the point $x=a$:
\begin{equation}
\varphi\left( a + \frac{y}{\beta p} \right) = \varphi(a)
+ \varphi'(a) \frac{y}{\beta p}  + \frac{1}{2!} \varphi''(a)
\left( \frac{y}{\beta p} \right)^2 + \cdots . 
\end{equation}
After simple algebra one thus arrives at the large-$\beta p$ expansion
\begin{equation}
l = a + \frac{1}{\beta p} - \frac{\beta \varphi'(a)}{(\beta p)^2}
+\frac{\left[ \beta\varphi'(a)\right]^2 - 2\beta \varphi''(a)}{(\beta p)^3}  
+ O \left( \frac{1}{(\beta p)^4} \right) .  
\end{equation}
Consequently,
\begin{eqnarray}
Z(\beta,n) & = & \frac{1}{1-an} -\beta \varphi'(a) a
-\beta \left[ \varphi'(a) a + 2 \varphi''(a) a^2 \right] (1-an)
\nonumber \\ & & + O\left[(1-an)^2\right] . \label{Znan} 
\end{eqnarray}
The leading singular term of this expansion is universal, i.e.,
independent of model's parameters, except of the hard core diameter $a$.
Note that, in contrast to the $T=0$ ground state when $l=a$ for all $p>p_i$,
an infinite pressure $p$ is necessary to compress particles to their
hard core $a$ if $T>0$.

\subsection{Tonks gas} \label{Sec2.4}
As a reference for comparison with further results, let us recall
the exact solution of the simplest 1D model of hard rods with
the interaction potential (\ref{hr}).

The Laplace transform of the pair Boltzmann factor (\ref{Om}) becomes
\begin{equation}
\widehat{\Omega}(s) = \frac{{\rm e}^{-a s}}{s} .
\end{equation}
Applying (\ref{recdensity}), the EoS for the reciprocal density reads as
\begin{equation} \label{EoShr}
l = a + \frac{T}{p}.
\end{equation}
For negligible $a\to 0$ (pointlike particles) or large temperatures,
one gets the EoS of the ideal gas $\beta p = n$.

In the low-temperature limit $T\to 0$ and for any positive pressure $p>0$,
$l(0)=a$ becomes the lattice constant of the close packed array of
hard rods in the ground state.

Increasing temperature $T$, the averaged distance between particles
$l$ grows linearly with $T$ and diverges in the high-temperature limit
$T\to\infty$.

Since it holds that
\begin{equation} 
\beta p = \frac{n}{1-a n} = n + a n^2 + a^2 n^3 +...,
\end{equation}
the virial coefficients are simply $B_k=a^{k-1}$ $(k=1,2,\ldots)$.
Note that for $d$-dimensional particles (hyperspheres) of diameter $a$,
the virial coefficients are expected to have the form $B_k=b_k(d) a^{k-1}$
\cite{Clisby05,Rohrmann08}, thus $b_k(1)=1$.

The compressibility factor $Z$ (\ref{Z}) is given by
\begin{equation} 
Z = \frac{1}{1-a n} .
\end{equation}

\renewcommand{\theequation}{3.\arabic{equation}}
\setcounter{equation}{0}

\section{Square shoulder model} \label{Sec3}
The square shoulder model is defined by the interaction potential
(\ref{general}) with \cite{Santos16}
\begin{equation} \label{SW}
\varphi(x) = \varepsilon , \qquad a < \vert x\vert < a' ,
\end{equation}
where $\varepsilon>0$ is a positive real parameter.
The length scale can be chosen so that the hard-core diameter $a=1$ and
the energy scale $\varepsilon=1$.
This potential is discontinuous at $\vert x\vert = a,a'$.

The Laplace transform (\ref{Om}) becomes
\begin{equation} \label{Shoulder}
\widehat\Omega(s) = \frac{\e^{-a' s}+\e^{-\beta \varepsilon}
(\e^{-a s}-\e^{-a' s})}{s}
\end{equation}
and the reciprocal density (\ref{recdensity}) reads as
\begin{equation} \label{Shoulderl}
l(T,p) = a' + \frac{T}{p} - \frac{a'-a}{1-\e^{-(a'-a)p/T}
+ \e^{[\varepsilon-(a'-a)p]/T}}.
\end{equation}

\begin{figure}[t]
\centering
\setbox1=\hbox{\includegraphics[clip,height=8.5cm]{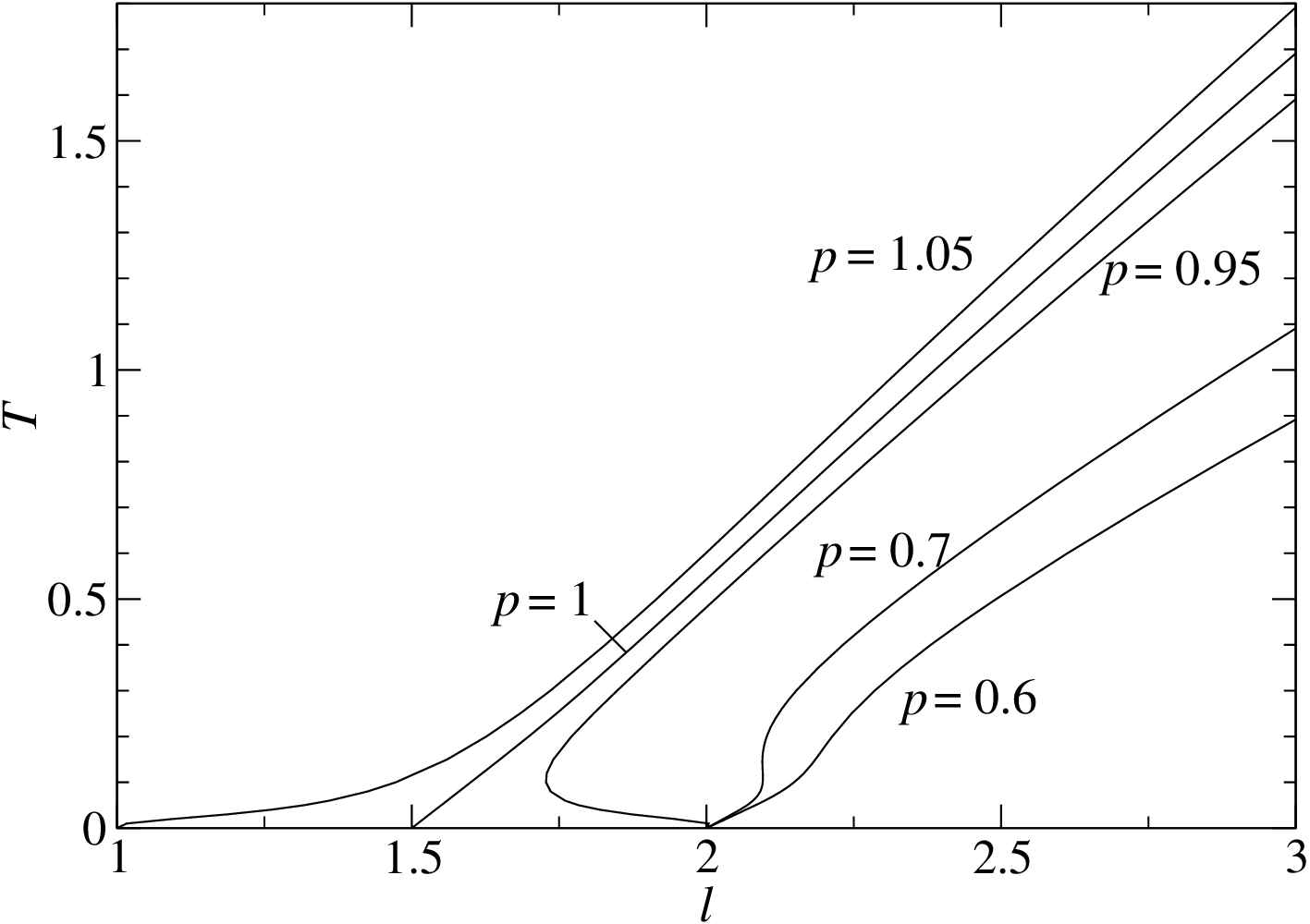}}
\includegraphics[clip,height=8.5cm]{fig1.eps}\lapbox[5cm]{-10.9cm}
{\raisebox{4.1cm}{\includegraphics[clip,height=3.9cm]{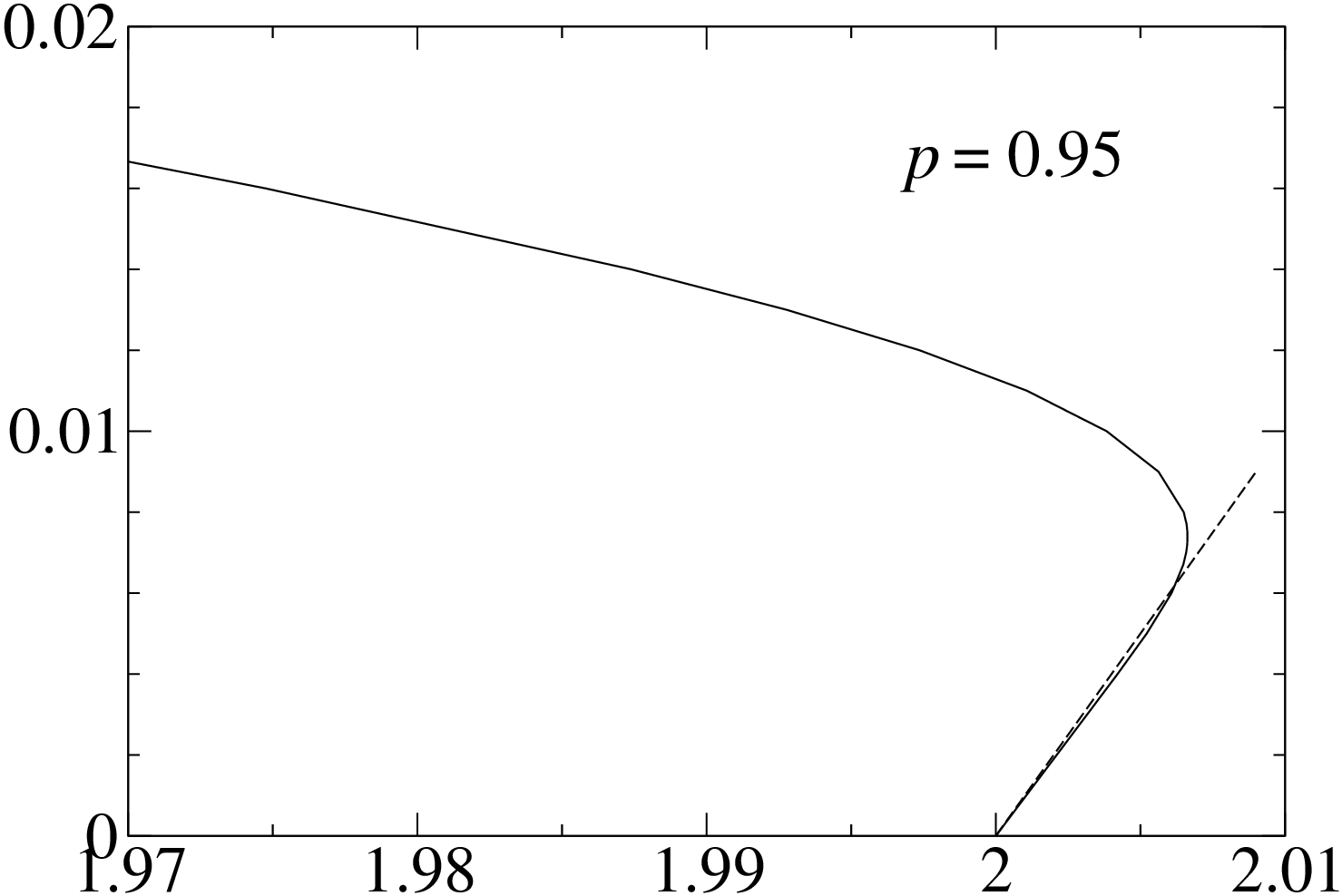}}}
\caption{Results for the square shoulder model
(\ref{SW}) with parameters $a=1$, $a'=2$ and $\varepsilon=1$ for
which the compressibility pressure $p_i=\varepsilon/(a'-a)=1$.
The dependence of $T$ versus $l$ is pictured for five values
of the pressure $p=1.05,1,0.95,0.7$ and $0.6$.
The inset magnifies the low-temperature region around $l=2$ for $p=0.95$.
See the text for a detailed description of the plots.}
\label{F1}
\end{figure}

The low-temperature analysis of the EoS (\ref{Shoulderl}) depends on the
signs of the arguments in the two exponentials.
Consequently, the low-$T$ expansion of $l(T,p)$ depends on the strength of $p$: 
\begin{equation} \label{shoulderzeroT}
l = \left\{
\begin{array}{ll}
\displaystyle{a + \frac{T}{p} + O(\e^{-[(a'-a)p-\varepsilon]/T})} &
\mbox{if $\displaystyle{p>\frac{\varepsilon}{a'-a}}$,} \cr
\displaystyle{\frac{a+a'}{2} + \frac{T}{p} + O(\e^{-(a'-a)p/T})} &
\mbox{if $\displaystyle{p=\frac{\varepsilon}{a'-a}}$,} \cr
\displaystyle{a' + \frac{T}{p} + O(\e^{-[\varepsilon-(a'-a)p]/T})} &
\mbox{if $\displaystyle{p<\frac{\varepsilon}{a'-a}}$.}  
\end{array} \right.   
\end{equation}
The incompressibility pressure above which $l=a$ is thus given by
$p_i=\varepsilon/(a'-a)$.

Let us first analyse the leading ground-state terms obtained in the
limit $T\to 0$.
In the region of large pressures $p>p_i$, the particles
form a close packed array with spacing $a$.
The ground state is degenerate at $p=p_i$ when all pairs of neighbouring
particles have spacing either $a$ or $a'$ (in one to one ratio)
and the mean spacing is $(a+a')/2$.   
The spacing skips to $a'$ when $p<p_i$.
Note that the prescription (\ref{derener}) does not apply
to the square shoulder potential (\ref{SW}).

Although the subleading term $T/p$ is the same for all values of $p>0$,
the next exponentially small corrections of type $\exp(-c/T)$ depend
on whether the pressure is above, at or below its compressibility value $p_i$.

Explicit calculations are made for the square shoulder model (\ref{SW})
with parameters $a=1$, $a'=2$ and $\varepsilon=1$
for which the compressibility pressure $p_i=\varepsilon/(a'-a)=1$.
The plots of $T$ versus $l$ calculated from the EoS (\ref{Shoulderl})
are pictured for five fixed values of the pressure
$p=1.05,1,0.95,0.7$ and $0.6$ in figure \ref{F1}.
The curve for $p=1.05$ (above $p_i$) starts from the ground state with
$l(0)=a=1$ and $l$ grows monotonously with increasing $T$ as expected;
the linear correction $T/p$ is visible in the region of low $T$.
The curve for the compressibility pressure $p_i=1$ starts from
the ground state with the mean $l(0)=(a+a')/2=3/2$ and $l$ grows monotonously
with $T$ as well.
The curve for $p=0.95$ (just below $p_i$) starts from the ground state
with $l(0)=a'=2$.
As is seen in the inset, $l$ grows like $T/p$ for very low $T$
(dashed line) but very soon, at around $T\approx 0.006$, $l$ becomes
a decreasing function of $T$ which is the NTE counterintuitive phenomenon.
This decay lasts inside a short interval of temperatures and subsequently
the expected growth of $l$ with increasing $T$ takes place up to $T\to\infty$.
The curve for $p=0.7$ behaves qualitatively as the one for $p=0.95$,
but the anomalous decay of $l$ with $T$ is less pronounced.
Finally, the curve for $p=0.6$ exhibits a monotonous growth of $l$
with $T$, in analogy with curves at and above $p_i=1$.

As is clear from figure \ref{F1}, the change from the anomalous NTE
decay of $l$ to its anticipated growth with increasing $T$ takes place
somewhere between $p=0.6$ and $p=0.7$.
Let this change occur at some temperature $T^*(a,a',\varepsilon)$
and pressure $p^*(a,a',\varepsilon)$.
The values of $T^*$ and $p^*$ are determined by two equations for
the inflection point:
\begin{equation} \label{inflection}
\frac{\partial l(T,p)}{\partial T}\Bigg\vert_{T^*,p^*} = 0 , \qquad
\frac{\partial^2 l(T,p)}{\partial T^2}\Bigg\vert_{T^*,p^*} = 0 .  
\end{equation}  
For the chosen parameters $a=1$, $a'=2$ and $\varepsilon=1$, and with
the use of formula (\ref{Shoulderl}) for $l(T,p)$, the solution
of these equations reads as
\begin{equation}
p^*(1,2,1) = 0.694472337\ldots , \quad
T^*(1,2,1) = 0.127904219\ldots ;
\end{equation}
we see that the value of $p^*(1,2,1)$ is indeed in between 0.6 and 0.7.  
To obtain $T^*$ and $p^*$ for arbitrary values of $a$, $a'$ and $\varepsilon$,
notice that formula (\ref{Shoulderl}) for $l(T,p)$ can be rewritten as follows
\begin{equation}
l(T,p;a,a',\varepsilon) = a' + (a'-a) \frac{\overline{T}}{\overline{p}}
\left[ 1 - \frac{\overline{p}}{\overline{T}}
\frac{1}{1-\e^{-\overline{p}/\overline{T}}+ \e^{(1-\overline{p})/{\overline T}}} \right] ,
\end{equation}
where
\begin{equation} \label{transform}
\overline{T} = \frac{1}{\varepsilon} T , \qquad
\overline{p} = \frac{a'-a}{\varepsilon} p .
\end{equation}  
Consequently,
\begin{equation}
l(T,p;a,a',\varepsilon) = (a'-a) l(\overline{T},\overline{p};1,2,1)
+ (2a-a') .  
\end{equation}  
Since the transformation (\ref{transform}) relates separately
temperatures and pressures, the inflection-point condition
(\ref{inflection}) for $l(T,p;a,a',\varepsilon)$ is equivalent
to the inflection-point condition (taken with $\overline{T}$-variable)
for $l(\overline{T},\overline{p};1,2,1)$.
Thus,
\begin{equation}
T^*(a,a',\varepsilon) = \varepsilon T^*(1,2,1) , \qquad
p^*(a,a',\varepsilon) = \frac{\varepsilon}{a'-a} p^*(1,2,1) .
\end{equation}

In the limit of high temperatures $T\to\infty$, the EoS (\ref{Shoulderl})
yields the expansion
\begin{equation} \label{EoSSHT}
l = \frac{1}{\beta p} + a + (a'-a)\varepsilon \beta
- \frac{a'-a}{2}\varepsilon[\varepsilon+2(a'-a)p] \beta^2 + O(\beta^3) .
\end{equation}
It can be checked that this expansion follows from the general formula
(\ref{HT}).

After simple algebra one obtains from the EoS (\ref{Shoulderl}) 
the virial expansion of type (\ref{Virhr}) with the coefficients
\begin{eqnarray} \label{SqWvir}
B_2 & = & a'-(a'-a) \e^{-\beta\varepsilon} , \nonumber \\
B_3 & = & a'^2 + (a-3a')(a'-a) \e^{-\beta\varepsilon}
+ 2(a'-a)^2 \e^{-2\beta\varepsilon} , \nonumber \\
B_4 & = & a'^3 - \frac{1}{2}(a'-a)(a^2-8a a'+13a'^2) \e^{-\beta\varepsilon}
\nonumber\\ & & + \frac{3}{2}(a'-a)^2(7a'-3a) \e^{-2\beta\varepsilon}
- 5(a'-a)^3 \e^{-3\beta\varepsilon} ,
\end{eqnarray}
etc.

\begin{figure}[t]
\begin{center}
\includegraphics[clip,width=0.84\textwidth]{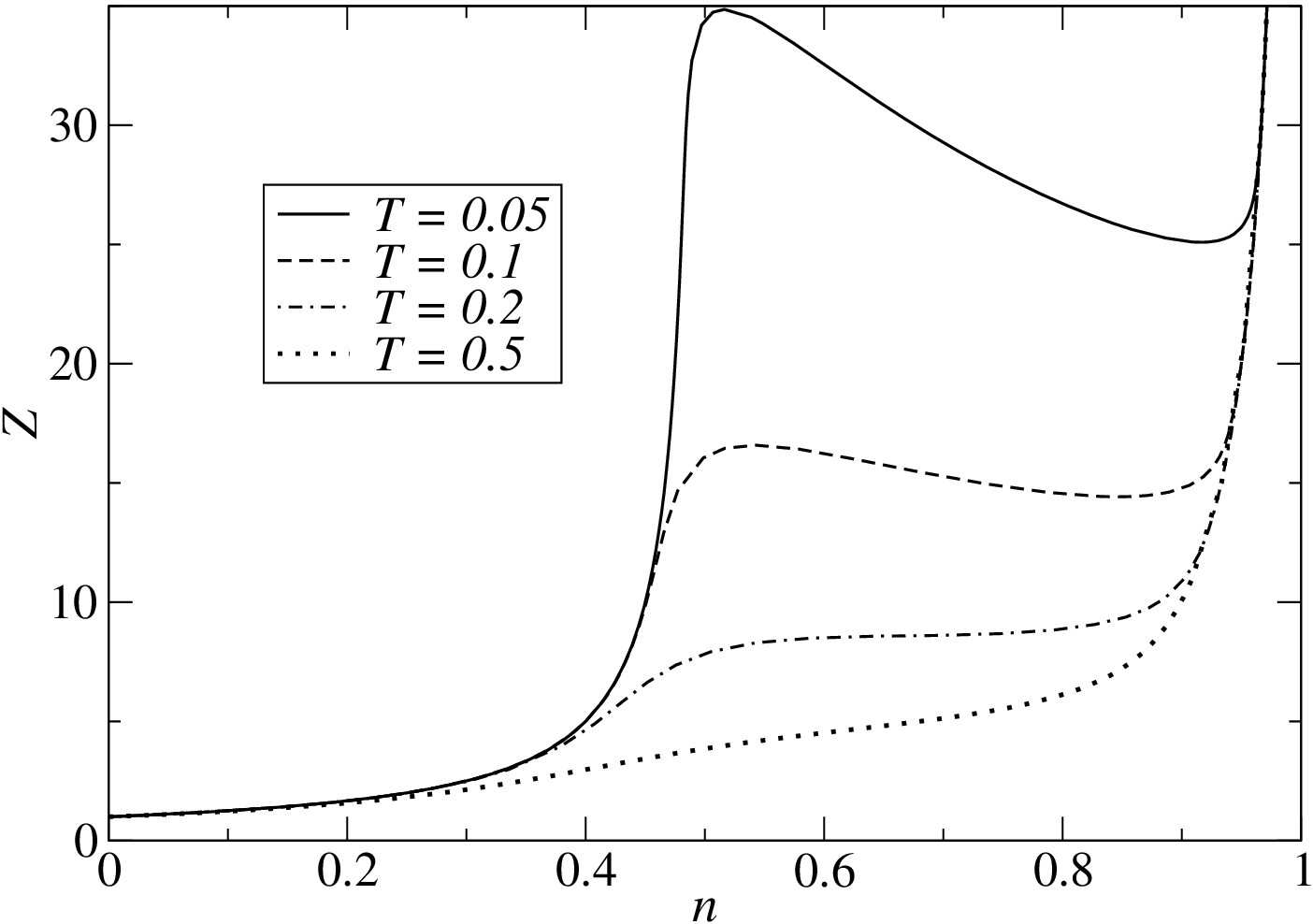}
\caption{The plot of the compressibility factor $Z$ versus
the particle density $n$ for the square shoulder potential (\ref{SW})
with parameters $a=1$, $a'=2$ and $\varepsilon=1$.
The isotherms $T=0.005, 0.1, 0.2, 0.5$ are represented by the solid,
dashed, dash-dotted and dotted curves, respectively.
See the text for a detailed description of the plots.}
\label{znsh}
\end{center}
\end{figure}

The plots of the compressibility factor $Z$ versus the particle density $n$
for different isotherms $T=0.05, 0.1, 0.2, 0.5$ is pictured in figure
\ref{znsh}.
The plots are nonmonotonic with the presence of
one maximum/minimum for low temperatures $T=0.05$ and $0.1$.
On the other hand, the plots of $Z(n)$ exhibit the monotonic growth
in the whole interval of $n\in [0,1]$ for higher temperatures
$T=0.2$ and $0.5$.
The curves collapse into the universal curve $1/(1-n)$
approaching the close packed limit $n\to 1^-$.
Since for the square shoulder potential (\ref{SW}) all derivatives
of $\varphi(x)$ at $x=0$ are equal to zero, corrections to the leading
term $1/(1-n)$ do not follow the formula (\ref{Znan}) but they
are proportional to the exponential $\exp[-(a'-a)\beta p]$
which is extremely small in the limit $\beta p\to\infty$.

\renewcommand{\theequation}{4.\arabic{equation}}
\setcounter{equation}{0}

\section{Ramp models} \label{Sec4}
The repulsive ramp models are defined by continuous decays from
$\phi(x=a^+)=\varepsilon>0$ to $\phi(a')=0$.
In this section, we present the explicit results for the linear, quadratic,
fusion of linear/quadratic and square-root ramp models.

\subsection{Linear ramp} \label{Sec4.1}
The interaction potential of the linear ramp model is defined by
(\ref{general}) with 
\begin{equation} \label{lrm}
\varphi(x) = \varepsilon \frac{(a'-\vert x\vert)}{(a'-a)} , \qquad
a< \vert x\vert < a'.
\end{equation}
It is continuous in the whole interval $(a,a']$, but its first derivative
with respect to $x$ is discontinuous at $\vert x\vert = a'$.
This core-softened pair potential describes water-like anomalies
\cite{Lomba07}.
Its formal 1D solution was discussed in \cite{Montero19}.

The Laplace transform (\ref{Om}) of the Boltzmann factor of
the potential (\ref{lrm}) reads as
\begin{equation} \label{EosRampl}
\widehat\Omega(s) = \frac{\beta\varepsilon\e^{-a's}-(a'-a)s
\e^{-\beta\varepsilon-as}}{s [\beta\varepsilon-(a'-a)s]} 
\end{equation}
and the reciprocal density (\ref{recdensity}) takes the form
\begin{equation} \label{Rampl}
l = a - \frac{(a'-a) T}{[\varepsilon-(a'-a)p]}
+ \frac{\varepsilon[T+(a'-a) p]}{
p\left\{\varepsilon-(a'-a)p \e^{-[\varepsilon-(a'-a)p]/T}\right\}}.
\end{equation}

\begin{figure}[t]
\begin{center}
\includegraphics[clip,width=0.84\textwidth]{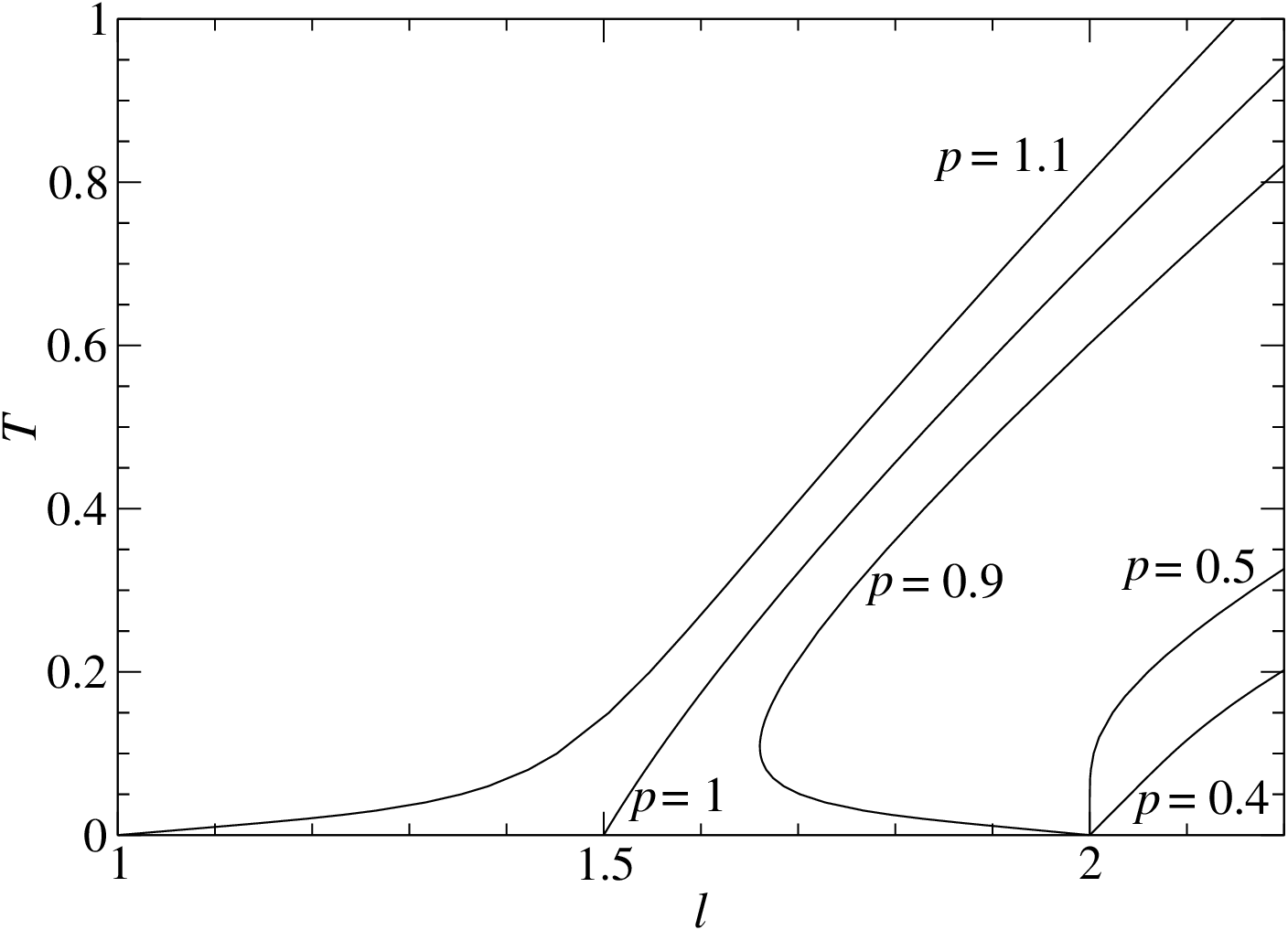}
\caption{Results for the linear ramp model (\ref{lrm})
with parameters $a=1$, $a'=2$, $\varepsilon=1$ and the compressibility
pressure $p_i=\varepsilon/(a'-a)=1$.
The dependence of $T$ versus $l$ is pictured for five fixed values
of the pressure $p=1.1,1,0.9,0.5$ and $0.4$.
See the text for a detailed description of the plots.}
\label{F2}
\end{center}
\end{figure}

The low-temperature expansion of the EoS (\ref{Rampl}) depends on
the sign of the argument of the exponential:
\begin{equation} \label{lr}
l = \left\{
\begin{array}{ll}
\displaystyle{a + \frac{(a'-a)T}{(a'-a)p-\varepsilon} +
O(\e^{-[(a'-a)p-\varepsilon]/T})} &
\mbox{if $\displaystyle{p>\frac{\varepsilon}{a'-a}}$,} \cr
\displaystyle{\frac{a+a'}{2} + \frac{T}{2p} + \frac{T^2}{2(a'-a)p^2}
+O(T^3)} &
\mbox{if $\displaystyle{p=\frac{\varepsilon}{a'-a}}$,} \cr
\displaystyle{a'+\frac{[\varepsilon-2(a'-a)p] T}{p[\varepsilon-(a'-a)p]}
+ O(\e^{-[\varepsilon-(a'-a)p]/T})} &
\mbox{if $\displaystyle{p<\frac{\varepsilon}{a'-a}}$.}  
\end{array} \right.   
\end{equation}
The incompressibility pressure is given by $p_i=\varepsilon/(a'-a)$.
Note that the prefactor to $T$ on the first line of (\ref{lr}) diverges when
$p\to p_i$ from above; to obtain the formula for
$p=p_i$ on the second line, one has to put
$p=p_i+\delta$, to expand the EoS (\ref{Rampl}) in powers of
small $\delta$ and to set $\delta=0$ at the end of the calculation.

In the ground-state, the system behaves similarly as the previous one
with the square shoulder potential: the particles form an
equidistant array with spacing $a$ ($a'$) for large $p>p_i$
(small $p<p_i$) pressures, with a coexistence point at $p=p_i$.
As before, the ground-state prescription (\ref{derener}) cannot be applied
to the present model.

Based on the formula $\lim_{\beta\to\infty} \widehat{\Omega}(s) = \e^{-a's}/s$
following from (\ref{EosRampl}), only the possibility of hard rods of
length $l=a'$ in the ground state was obtained in Ref. \cite{Montero19}.
The problem is that $l$ is determined by $\widehat{\Omega}(s)$ at
$s=\beta p$, the infinite $\beta$ entering also the argument of
$\widehat{\Omega}(s)$ which was ignored previously.
Hard rods of length $a'$ thus correspond to the ground state only in
the limit $p\to 0$, leaving $s=\beta p$ finite, in agreement
with our findings.

The subleading terms are always linear in $T$, but the prefactor depends on
the region of the pressure.
The prefactor is positive for $p>p_i$, going to $\infty$
when $p$ approaches its compressibility value $p_i$ from above.
It is positive and finite at $p=p_i$.
As concerns the region of $0<p<p_i$, there is a special value of
$p_0(a,a',\varepsilon)=\varepsilon/[2(a'-a)]$
(with the zero prefactor) dividing this pressure region onto two halves:
the prefactor is negative if $p_0<p<p_i$, signalizing
the anomalous low-$T$ behavior, and positive if $0<p<p_0$, signalizing
a normal increase of $l$ with $T$ (at least in the
low-$T$ region).
The next correction terms are exponentially small, of type $\exp(-c/T)$,
except for the transition pressure $p=\varepsilon/(a'-a)$ when $l$
exhibits an analytic expansion in integer powers of $T$.

For the choice of the parameters $a=1$, $a'=2$ and $\varepsilon=1$,
the compressibility pressure $p_i=\varepsilon/(a'-a)=1$.
The plots of $T$ versus $l$, obtained by using the EoS
(\ref{Rampl}) for five fixed values of the pressure
$p=1.1,1,0.9,0.5$ and $0.4$, are presented in figure \ref{F2}.
These values of $p$ are chosen such that the discussion follows the previous
one for the square shoulder model in figure \ref{F1} with one exception:
the tangent to the curves does not change its sign for very low $T$
as it was pictured in the inset of figure \ref{F1}.

The high-temperature expansion of $l$ is rather simple as usual
\begin{equation} \label{EoSRLHT}
l = \frac{1}{\beta p} + a + \frac{(a'-a)\varepsilon}{2}\beta
-\frac{a'-a}{6}\varepsilon[\varepsilon+2(a'-a)p] \beta^2 + O(\beta^3) .
\end{equation}

The coefficients of the virial expansion (\ref{Virhr}) become
\begin{eqnarray} \label{Ramplvir}
B_{2}&=&a'-\frac{(a'-a)(1-\e^{-\beta\varepsilon})}{\beta\varepsilon}\nonumber\\
B_{3}&=&\frac{2(a'-a)^2\e^{-2\beta\varepsilon}
-2(a'-a)\e^{-\beta\varepsilon}[a'-a-(2a'-a)\beta\varepsilon]}{\beta^2\varepsilon^2}
\nonumber\\ &&
+ \frac{a'\beta\varepsilon[2a+a'(\beta\varepsilon-2)]}{\beta^2\varepsilon^2} ,
\end{eqnarray}
etc.

The plots of the compressibility factor $Z$ versus
the particle density $n$ at fixed temperature $T$ are similar to those of
the square shoulder potential in figure \ref{znsh}.
Namely, for low temperatures the plots of $Z(n)$ are nonmonotonic with
one maximum/minimum and for higher temperatures the plots of $Z(n)$
exhibit the monotonic growth in the whole interval of $n\in [0,1]$.
The plot of $Z(n)$ is not presented here to save space.

\subsection{Quadratic ramp} \label{Sec4.2}
Being motivated by a standard model of an equidistant array of harmonically
coupled particles vibrating around their ground-state positions,  
let us consider the ramp model with the quadratic
interaction potential \cite{Percus87}
\begin{equation} \label{Q}
\varphi(x) = \varepsilon\frac{(a'-\vert x\vert)^2}{(a'-a)^2} , \qquad
a< \vert x\vert < a' .
\end{equation}
The parameters are chosen so that the potential $\varphi(x)$ and its first
derivative $\varphi'(x)$ are continuous (namely they vanish) at $x=a'$.

The Laplace transform (\ref{Om}) of the Boltzmann factor of
the potential (\ref{Q}) takes the form
\begin{eqnarray} \label{QOm}
\hat\Omega(s) & = & \frac{\e^{-a' s}}{s} +
\frac{(a'-a)\exp{\left[ -a's +\frac{(a'-a)^2 s^2}{4\beta \varepsilon}\right]}}{
2\sqrt{\beta \varepsilon}} \sqrt{\pi} \nonumber\\ & & \times
\left\{ {\rm erf}\left[\frac{(a'-a)s}{2\sqrt{\beta \varepsilon}}\right]
+{\rm erf}\left[\frac{2\beta \varepsilon-(a'-a)s}{2\sqrt{\beta \varepsilon}}
\right]\right\} ,
\end{eqnarray}
where
\begin{equation} \label{error}
{\rm erf}(z) = \frac{2}{\sqrt{\pi}} \int_0^z {\rm d}t\, \e^{-t^2}
\end{equation}
is the error function \cite{Gradshteyn}.
Since the formula for the reciprocal density (\ref{recdensity}) is rather
complicated, it is not presented here.

\begin{figure}[t]
\begin{center}
\includegraphics[clip,width=0.84\textwidth]{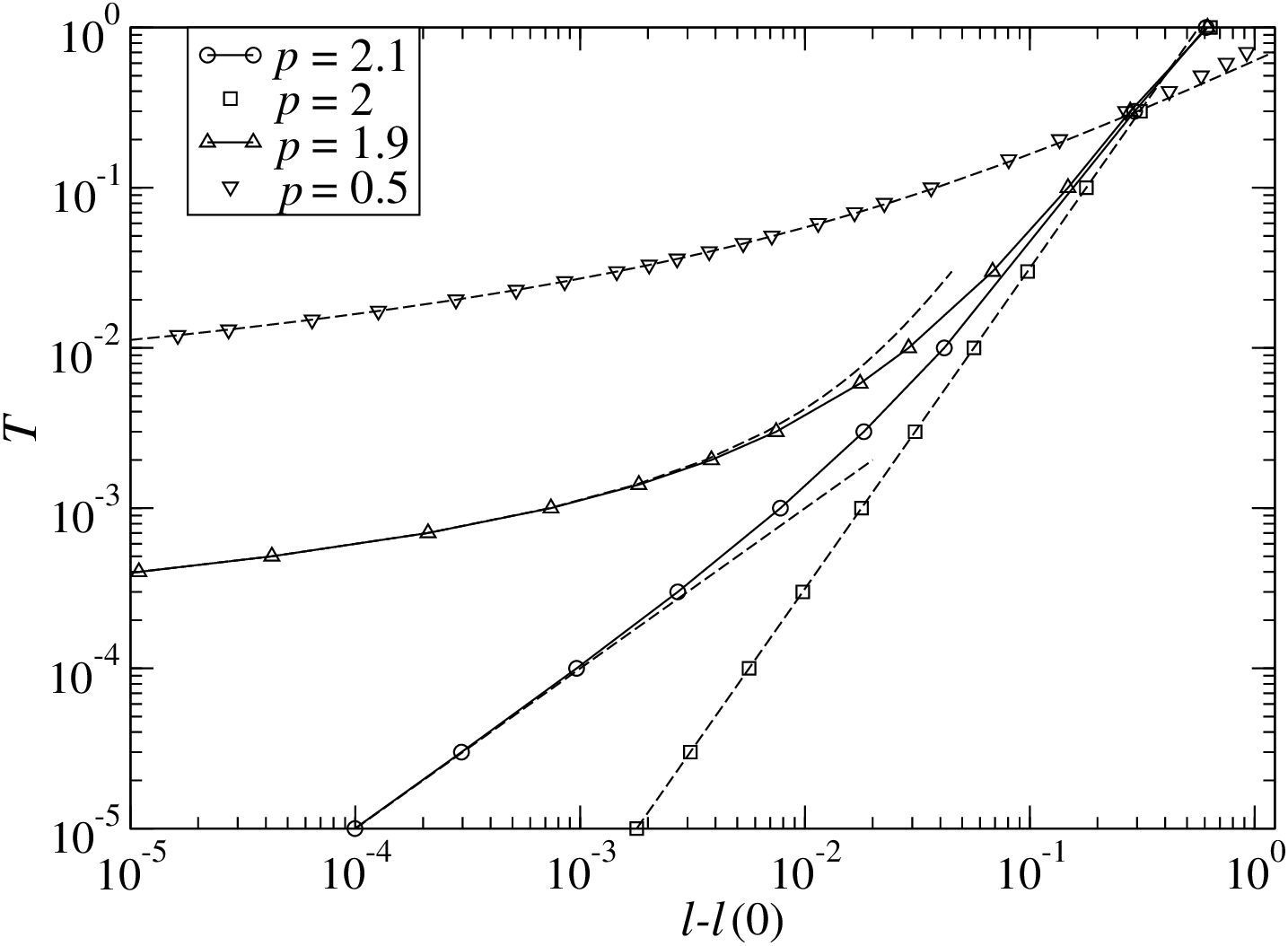}
\caption{The plots of $T$ versus $l-l(0)$ in the log-log scale for
the quadratic-ramp potential (\ref{Q}) with the parameters $a=1$, $a'=2$
and $\varepsilon=1$, for four fixed values of the pressure
$p=2.1,2,1.9,0.5$ (open symbols).
Dashed curves depict the subleading terms in (\ref{lQ}).}
\label{F3}
\end{center}
\end{figure}

The derivative $\partial\varphi(x)/\partial x$ is continuous for
$a<x\le a'$ and it vanishes for both $x\to {a'}^-$ and $x\to {a'}^+$.
The EoS for this interaction potential at $T=0$ can be thus deduced from
the ground-state formula (\ref{derener}), $p=2\varepsilon[a'-l(0)]/(a'-a)^2$.
The spacing of the equidistant array of particles $l(T=0)$ depends
continuously on $p$ as follows
\begin{equation}
l(0) = \left\{
\begin{array}{ll}  
\displaystyle{a} & \mbox{if $\displaystyle{p\ge\frac{2\varepsilon}{a'-a}}$,} \cr
\displaystyle{a'- \frac{p(a'-a)^2}{2\varepsilon}} & \mbox{if
$\displaystyle{p<\frac{2\varepsilon}{a'-a}}$.}
\end{array} \right.
\end{equation}  
The incompressibility pressure is given by $p_i=2\varepsilon/(a'-a)$.

The leading term of the low-temperature expansion of $l(T,p)-l(0,p)$
takes the form
\begin{equation} \label{lQ}
\hspace{-57pt} l - l(0) = \left\{
\begin{array}{ll}
\displaystyle{\frac{(a'-a)}{q}T-\frac{4(a'-a)\varepsilon }{q^3}T^2
+ O(T^3)} & \mbox{if $\displaystyle{p>\frac{2\varepsilon}{a'-a}}$,} \cr
\displaystyle{\sqrt{\frac{2(a'-a)}{\pi p}}\sqrt{T}
+ O\left( T^{3/2}\e^{-\frac{(a'-a) p}{2 T}}\right)} &
\mbox{if $\displaystyle{p=\frac{2\varepsilon}{a'-a}}$,} \cr
\displaystyle{\frac{a'-a}{2}\sqrt{\frac{T}{\pi\varepsilon}}
\e^{-\frac{[2\varepsilon-(a'-a)p]^2}{4\varepsilon T}}+\cdots} & \mbox{if
$\displaystyle{\frac{\varepsilon}{a'-a}}\le p < \frac{2\varepsilon}{a'-a}$,}
\cr \displaystyle{\frac{1}{(a'-a)p^2}\sqrt{\frac{\varepsilon T^3}{\pi}}
\e^{-\frac{(a'-a)^2p^2}{4\varepsilon T}}+\cdots} &
\mbox{if $\displaystyle{p<\frac{\varepsilon}{a'-a}}$.}
\end{array} \right.
\end{equation}
Here, the parameter $q=p(a'-a)-2\varepsilon$ was introduced.
Note that all leading terms of the $T$-expansion are such that $l-l(0)$ grows
with increasing $T$ which is a sign of the expected behaviour, at least
in the region of low $T$.

For the usual choice of the parameters $a=1$, $a'=2$ and $\varepsilon=1$,
the plots of $T$ versus $l-l(0)$ for $p=2.1,2,1.9$ and $0.5$
are pictured in the log-log scale in figure \ref{F3}.
The symbolic language {\em Mathematica} was used to calculate
values plotted in all figures in this paper and usually it is enough to use
20 digits precision, but for the exponentially small values of $l(T)-l(0)$
with $T<10^{-3}$ and $p<2$ the accuracy up to 200 digits is necessary.
The asymptotics written in (\ref{lQ}) are depicted by dashed curves which fit
with the plots over a large interval of $T$-values.
The lower two dashed lines are the linear asymptotes with the slopes 2 and 1
for $p=p_i=2$ (open squares) and $p=2.1>p_i$ (open circles), respectively.
The other two dashed curves (open triangles) $p=1.9<p_i$
and $p=0.5<\varepsilon/(a'-a)$ correspond to exponential corrections of
type $\sqrt{T}\exp(-\alpha/T)$ and $T^{3/2}\exp(-\alpha/T)$, respectively.
It is seen that for every value of the pressure $p$ that the averaged lattice
spacing $l$ is the monotonously increasing function of $T$.

The high-temperature expansion of $l$ is of simple form
\begin{equation} \label{EoSQHT}
l = \frac{1}{\beta p} + a + \frac{(a'-a)\varepsilon}{3}\beta
- \frac{a'-a}{30}\varepsilon[3\varepsilon+5(a'-a)p] \beta^2 + O(\beta^3) .
\end{equation}
The large-$l$ behavior $T(l)$ is given by $T\approx p l-p a+O(1/T)$. 

The virial coefficients read as
\begin{eqnarray} \label{Qvir}
B_2 &=& a'+\frac{a-a'}{2}\sqrt{\frac{\pi}{\beta\varepsilon}}
{\rm erf}({\sqrt{\beta \varepsilon}}) \nonumber\\
B_3 &=& a'^2-\frac{(a'-a)^2}{\beta \varepsilon}(1-\e^{-\beta\varepsilon})
-\frac{(a'-a)a'\sqrt{\pi}}{\sqrt{\beta \varepsilon}}
{\rm erf}({\sqrt{\beta \varepsilon}})\nonumber\\
& & +\frac{(a'-a)^2\pi}{2\beta \varepsilon}
{\rm erf}^2({\sqrt{\beta \varepsilon}}) ,
\end{eqnarray}
etc.

For any $T$, the plot of $Z(n)$ exhibits the monotonic growth
in the whole interval of $n\in [0,1]$.

\subsection{Fusion of linear and quadratic ramps} \label{Sec4.3}
Next we propose a fusion of the above linear and quadratic ramps.
Taking $a=1$ and $a'=2$, the potential parameters are chosen such that
the linear and quadratic branches merge in the middle of the interval
$[a,a']$, at $(a+a')/2=3/2$, and the potential as well as its first
derivative are continuous in the whole interval $[1,2]$ and equal
to 0 at $x=a'$: 
\begin{equation}\label{lqphi}
\varphi(x) = \left\{
\begin{array}{ll}  
\displaystyle{-\frac{4}{3}x+\frac{7}{3}} &
\mbox{if $\displaystyle{1<x\le \frac{3}{2}}$,} \cr
\displaystyle{\frac{4}{3}(x-2)^2} & \mbox{if
$\displaystyle{\frac{3}{2}\le x\le 2}$.} \end{array} \right.
\end{equation}
Note that the second derivative $\varphi''(x)$ is discontinuous
at $x=3/2$, so that the function $\varphi(x)$ is not analytic at this point.

By using formula (\ref{Om}), one gets
\begin{eqnarray} 
\hat\Omega(s) & = & \frac{\e^{-2s}}{s} +
\frac{3\e^{-5s/2}\left(\e^{s-\beta/3}-\e^{3s/2-\beta}\right)}{4\beta -3s}
\nonumber \\ & & +
\frac{\sqrt{3\pi}\e^{3s^2/(16\beta)-2s}}{4\sqrt{\beta}}
\left[ {\rm erf}\left(\frac{\sqrt{3}s}{4\sqrt{\beta}}\right)
-{\rm erf}\left(\frac{3s-4\beta }{4\sqrt{3\beta}}
\right)\right]. \label{LQOm}
\end{eqnarray}
The formula for the reciprocal density $l$ (\ref{recdensity}) is
straightforward, but not presented here to save space. 
The low-temperature series of $l(T,p)$ reads as
\begin{equation} \label{lfusion}
l = \left\{
\begin{array}{ll}  
\displaystyle{1+\frac{T}{p-p_i}}+\cdots &
\mbox{if $\displaystyle{p\ge\frac{4}{3}}$,} \cr
\displaystyle{\frac{5}{4}+\frac{\sqrt{3\pi T}}{8}}+O(T)
& \mbox{if $\displaystyle{p=\frac{4}{3}}$,} \cr
\displaystyle{2-\frac{3}{8}p}-\cdots & \mbox{if
$\displaystyle{p<\frac{4}{3}}$.} \end{array} \right.
\end{equation} 
The value of the incompressibility pressure $p_i=4/3$ is related to
the slope of the linear part of $\varphi(x)$ and the dots denote
positive exponentially small terms of type $\exp(-c/T)$.

In the ground state, $l(0,p)$ jumps from $3/2$ at $p\to 4/3^-$
to $1$ at $p\to 4/3^+$ via the intermediate value $5/4=(1+3/2)/2$
at $p_i=4/3$.
The expression for $l(0,p)=2-3p/8$ for $p<4/3$ fulfills
the prescription (\ref{derener}) applied to the quadratic part of
the potential (\ref{lqphi}).

\begin{figure}[t]
\begin{center}
\includegraphics[clip,width=0.84\textwidth]{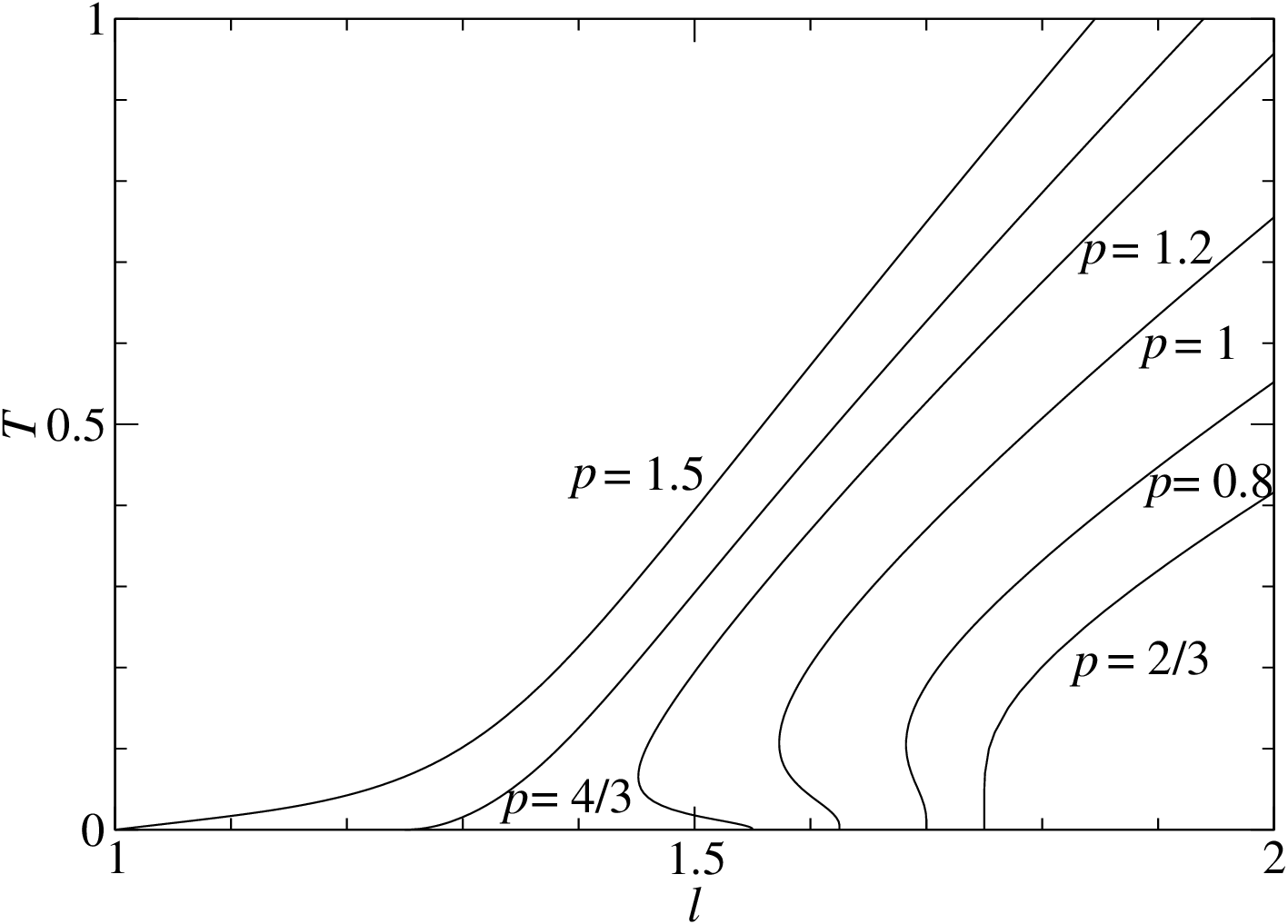}
\caption{The dependence of $T$ versus $l$ for
the fusion potential (\ref{lqphi}), for six values of the pressure
$p=1.5,4/3,1.2,1,0.8$ and $2/3$.
See the text for description of the plots.}
\label{Fig}
\end{center}
\end{figure}

The $T(l)$ dependence is plotted for six values of the pressure
$p=1.5,4/3,1.2,1,0.8$ and $2/3$ in figure \ref{Fig}.
The low-temperature NTE anomaly is present for the pressures $2/3<p<4/3$;
the relevant information is that in formula (\ref{lfusion}) for $l(T,p)$
the leading correction to the ground state for $p<4/3$ has the minus sign.

The high-temperature expansion
\begin{equation} \label{EoSLQHT}
l = \frac{1}{\beta p} + 1 + \frac{7}{18}\beta
- \left(\frac{17}{135}+\frac{5p}{24}\right) \beta^2 + O(\beta^3)
\end{equation}
can also be obtained with the aid of the general formula (\ref{HT}).

The first nontrivial virial coefficient is 
\begin{equation} \label{LQvir}
B_2 = 2+\frac{\e^{-\beta}}{4\beta}
\left(3-3\e^{-2\beta/3}\right)-\sqrt{\frac{3\pi}{16\beta}} \,
{\rm erf}\left( {\sqrt{\frac{\beta}{3}}}\right) .
\end{equation}

In analogy with the square shoulder and linear ramp
potentials, for low temperatures the plots of $Z(n)$ are nonmonotonic
with the presence of one maximum/minimum while for higher temperatures
$Z(n)$ exhibits the monotonic growth in the whole interval of $n\in [0,1]$.

\subsection{Ramp containing square root of distance} \label{Sec4.4}
There exist interaction potentials of the ramp type which are functions of
the square root of the distance, like
\begin{equation} \label{QSR1}
\varphi(x) = \varepsilon
\frac{\sqrt{a'}-\sqrt{\vert x\vert}}{\sqrt{a'}-\sqrt{a}} ,
\qquad a <\vert x \vert <a'
\end{equation}
or
\begin{equation} \label{QSR2}
\varphi(x) = \varepsilon
\left( \frac{\sqrt{a'}-\sqrt{\vert x\vert}}{\sqrt{a'}-\sqrt{a}} \right)^2 ,
\qquad a <\vert x \vert <a' . 
\end{equation}
Both potentials are continuous (they vanish) at $x=a'$.
While the derivative of the potential (\ref{QSR1}) with respect to $x$
is discontinuous at $x=a'$, the one of (\ref{QSR2}) is continuous at $x=a'$.

For both models, {\em Mathematica} yields quite complicated expressions
for $\widehat{\Omega}(s)$ and $l(T,p)$ in terms of the error function
(\ref{error}).
We do not write down cumbersome exact formulas, but rather comment
series expansions and results obtained from them. 

Results for the interaction potential (\ref{QSR1}) are such
that in certain intervals of $p$ the average distance between
particles $l$ as the function of $T$ exhibits the anomalous NTE decay
in the low-$T$ region.
Based on the previous results, this is intuitively associated with the fact
that $\varphi'(x)$ is discontinuous at $x=a'$.

\begin{figure}[t]
\begin{center}
\includegraphics[clip,width=0.84\textwidth]{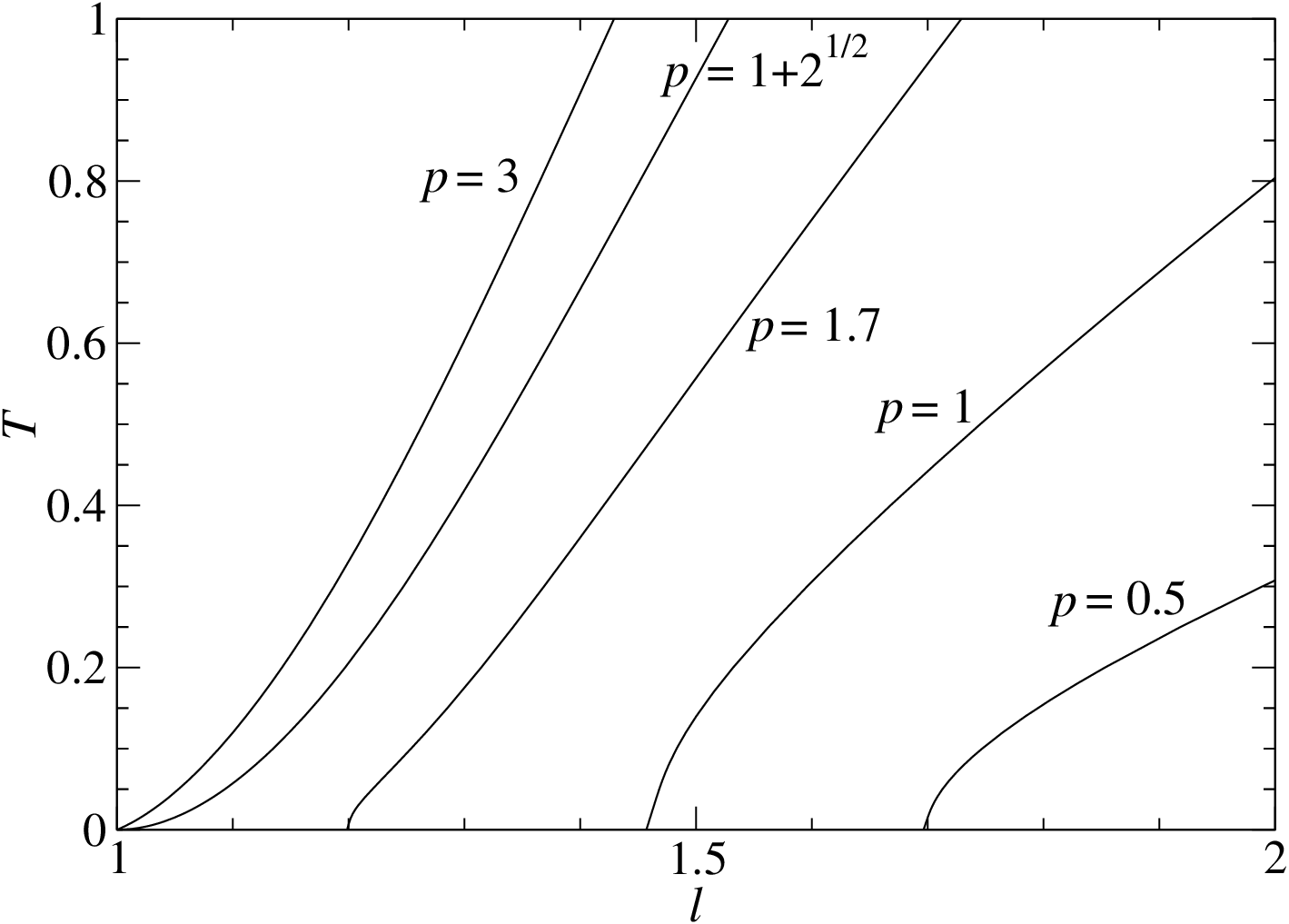}
\caption{Results for the square-root potential
(\ref{QSR2}) with parameters $a=1$, $a'=2$ and $\varepsilon=1$.
The dependence of $T$ versus $l$ is pictured for five values
of the pressure $p=3,1+\sqrt{2},1.7,1$ and $0.5$.
The compressibility value of $p_i=\varepsilon/(\sqrt{a a'}-a)=1+\sqrt{2}$
is characterized by the leading correction of type $\sqrt{T}$ in
(\ref{lQSRLT}).}
\label{F4}
\end{center}
\end{figure}

The potential (\ref{QSR2}) exhibits a more interesting behaviour. 
The ground-state results for the array spacing
\begin{equation} \label{l0sqr}
l(0) = \left\{
\begin{array}{ll}
a & \mbox{if $\displaystyle{p\ge \frac{\varepsilon}{\sqrt{a'a}-a}}$,} \cr
\displaystyle{\frac{a'\varepsilon^2}{[\varepsilon+(\sqrt{a'}-\sqrt{a})^2p]^2}}
& \mbox{if $\displaystyle{p<\frac{\varepsilon}{\sqrt{a'a}-a}}$}
\end{array} \right.  
\end{equation}  
can be also obtained directly by using the ground-state
prescription (\ref{derener}).
The incompressibility pressure is given by $p_i=\varepsilon/(\sqrt{a'a}-a)$.

The leading terms of the low-temperature expansion of $l(T,p)-l(0)$ read as
\begin{equation} \label{lQSRLT}
l-l(0) = \left\{
\begin{array}{ll}
\displaystyle{\frac{\sqrt{a'a}-a}{(\sqrt{a'a}-a)p-\varepsilon}T + O(T^{3/2})}&
\mbox{if $\displaystyle{p>\frac{\varepsilon}{\sqrt{a'a}-a}}$,} \cr
\displaystyle{2\frac{\sqrt{a'a}-a}{\sqrt{\varepsilon \pi\sqrt{a'/a}}}
\sqrt{T} + O(T)} &
\mbox{if $\displaystyle{p=\frac{\varepsilon}{\sqrt{a'a}-a}}$,} \cr
\displaystyle{\frac{3}{2} \frac{(\sqrt{a'}-\sqrt{a})^2}{
\varepsilon+(\sqrt{a'}-\sqrt{a})^2 p}T + \cdots} &
\mbox{if $\displaystyle{p<\frac{\varepsilon}{\sqrt{a'a}-a}}$,}
\end{array} \right.
\end{equation}
where dots denote an exponentially small term of type $\exp(-\alpha/T)$.
Note that the prefactor to $T$ on the first line diverges
as $p\to\varepsilon/(\sqrt{a'a}-a)$ from above.
As is evident from (\ref{lQSRLT}), all prefactors to $T$ and $\sqrt{T}$
are positive, i.e., $l(p,T)$ is an increasing function of $T$ in the region
of low $T$, regardless of the chosen $p$ and model's parameters. 
Exact results presented in figure \ref{F4} show that $l(p,T)$
is an increasing function of $T$ for {\em all temperatures}.

The high-temperature expansion of $l$ reads as
\begin{eqnarray} 
l & = & \frac{1}{\beta p} + a + \frac{\varepsilon}{6}(a'+2\sqrt{a'a}-3a)
\beta - \frac{\sqrt{a'}-\sqrt{a}}{30} \varepsilon
\left\{ \sqrt{a'}\left[ \varepsilon+2(a'+a)p\right] \right. \nonumber \\ & &
\left. + \sqrt{a}(5\varepsilon+6a'p)
-10 a^{3/2}p \right\} \beta^2 + O(\beta^3) . \label{lQSRHT}
\end{eqnarray}

The second virial coefficient becomes
\begin{eqnarray} \label{QSRvir}
B_2 &=& a'+(\sqrt{a'}-\sqrt{a})^2\frac{1-\e^{-\beta\varepsilon}}{\beta\varepsilon}
-(a'-\sqrt{a'a})\sqrt{\frac{\pi}{\beta\varepsilon}}
{\rm erf}\left( {\sqrt{\beta \varepsilon}}\right) .
\end{eqnarray}

For any $T$, the plots of the compressibility factor
$Z$ versus the particle density $n$ exhibit the monotonic growth in
the whole interval of $n\in [0,1]$.

\renewcommand{\theequation}{5.\arabic{equation}}
\setcounter{equation}{0}

\section{Logarithmic models} \label{Sec5}
In this section, we study three cases of the logarithmic interaction potential
which admit analytic solutions.

\subsection{Logarithmic potential of the ramp type} \label{Sec5.1}
Let us consider the ramp model with the logarithmic interaction potential
\begin{equation} \label{Ln}
\varphi(x) = -A\ln\left( \frac{\vert x\vert}{a'}\right) ,
\qquad a< \vert x\vert < a' ,
\end{equation}
where the real parameter $A>0$.
The potential has a finite (positive) value $\varepsilon=A\ln(a'/a)$
at $x=a$ and vanishes, but its derivative is discontinuous, at $x=a'$.

The Laplace transform (\ref{Om}) for the potential (\ref{Ln}) reads as
\begin{equation} \label{LnOm}
\widehat{\Omega}(s) = \frac{\e^{-a' s}}{s}+\frac{(a's)^{-A\beta}}{s}
\left[ \Gamma(1+A\beta,a s)-\Gamma(1+A\beta,a' s) \right] ,
\end{equation}
where $\Gamma(k,z)=\int_z^\infty t^{k-1}\e^{-t}{\rm d}t$ is the upper incomplete
gamma function \cite{Gradshteyn}.
The EoS takes the form 
\begin{eqnarray} 
\hspace{-70pt}l = \frac{\e^{(a'-a)\beta p}(a\beta p)^{1+A\beta}
+(a'\beta p)^{A\beta}+(1+A\beta)\e^{a'\beta p}
\left[ \Gamma(1+A\beta,a\beta p)-\Gamma(1+A\beta,a'\beta p)\right]}{{\beta p}
\left\{ (a'\beta p)^{A\beta}+\e^{a'\beta p} \left[ \Gamma(1+A\beta,a\beta p)-
\Gamma(1+A\beta,a'\beta p) \right]\right\}} . \nonumber \\ \label{tt}
\end{eqnarray}

\begin{figure}[t]
\begin{center}
\includegraphics[clip,width=0.84\textwidth]{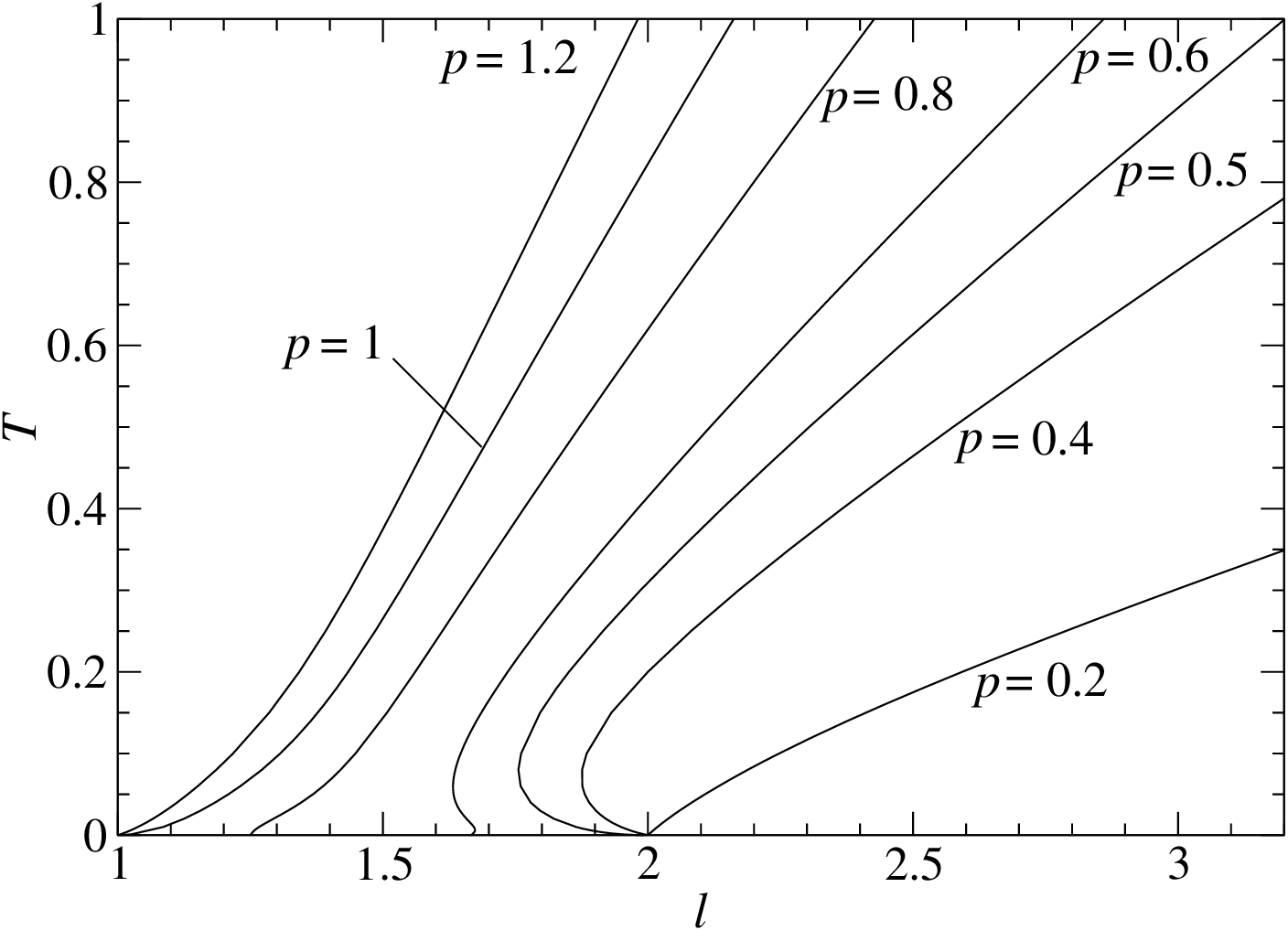}
\caption{The plot of $T$ versus $l$ for the logarithmic ramp model (\ref{Ln})
with the parameters $a=1$, $a'=2$ and $A=1$, for seven values of
the pressure $p=1.2,1,0.8,0.6,0.5,0.4,0.2$.
See the text for a detailed description of the plots.}
\label{F5}
\end{center}
\end{figure}

The low-temperature series expansion of $l$ splits into five possibilities:
\begin{equation} \label{lLn}
l =  \left\{
\begin{array}{ll}  
\displaystyle{a+\frac{T}{p-A/a} - \frac{2A T^2}{a^2(p-A/a)^3} + O(T^3)} &
\mbox{if $\displaystyle{p > \frac{A}{a}}$,} \cr
\displaystyle{a + a\sqrt{\frac{2}{\pi A}}\sqrt{T} + O(T)} &
\mbox{if $\displaystyle{p=\frac{A}{a}}$,} \cr
\displaystyle{\frac{A}{p}+\frac{T}{p}+\left( \frac{A}{p}+\frac{T}{p} \right)
\e^{-a p/T}+\cdots} & \mbox{if $\displaystyle{\frac{A}{a'}<p<\frac{A}{a}}$,} \cr
\displaystyle{a' - a' \sqrt{\frac{2}{\pi A}} \sqrt{T} + O(T)} &
\mbox{if $\displaystyle{p=\frac{A}{a'}}$,} \cr
\displaystyle{a' + \left( \frac{1}{p}+\frac{1}{p-A/a'} \right) T + O(T^2)} &
\mbox{if $\displaystyle{p<\frac{A}{a'}}$.}
\end{array} \right.
\end{equation}
The incompressibility pressure is given by $p_i=A/a$.

As concerns the ground state, the particles are pushed to $l=a$ for
large $p\ge p_i$.  
For medium pressures $A/a' < p < p_i$, the chain spacing $l = A/p$
interpolates continuously between $a$ and $a'$ and the formula
(\ref{derener}) does apply here.
The spacing $l=a'$ in the region of weak pressures $p\le A/a'$. 

The subleading terms are of order $T$, except for the pressures
$p=A/a$ and $p=A/a'$ when they are of order $\sqrt{T}$.  
The prefactor to $T$ diverges when $p\to A/a$ from above or
$p\to A/a'$ from below.
The prefactor to $T$ or $\sqrt{T}$ is positive for $p>A/a'$ and $p<A/(2a')$,
while it is negative for $A/(2a')<p\le A/a'$ which is a sign of the
anomalous NTE behaviour of $l(T)$ in this interval of pressures.

Calculations are made for the logarithmic ramp (\ref{Ln})
with parameters $a=1$, $a'=2$ and $A=1$, so that the important ratios
$A/a=1$, $A/a'=1/2$ and $A/(2a')=1/4$.
The plots of $T$ versus $l$, obtained by treating the EoS
(\ref{tt}), are pictured for seven fixed values of the pressure
$p=1.2,1,0.8,0.6,0.5,0.4,0.2$ in figure \ref{F5}.
The curves for $p=1.2$ and $p=1$ start from the ground state with
$l(0)=a=1$ and $l$ grows monotonously with increasing $T$.
The curve for $p=0.8$ starts from $A/p=5/4$ and $l$ grows monotonously
with $T$ as well.
The anomalous decay of $l$ with increasing $T$ occurs at the pressure
$p^*(a,a',A)$ and temperature $T^*(a,a',A)$ which are determined by
two equations for the inflection point (\ref{inflection}):
\begin{equation}
p^*(1,2,1) = 0.647693004\ldots, \quad
T^*(1,2,1) = 0.024628083\ldots .
\end{equation}
For arbitrary values of $a$, $a'$ and $\varepsilon$, following the procedure
in section \ref{Sec3} one gets
\begin{equation} \label{other}
p^*(a,a',A) = \frac{A}{a'-a} p^*(1,2,1) , \qquad
T^*(a,a',A) = A T^*(1,2,1) .
\end{equation}
As an example, the curve for $p=0.6<p^*(1,2,1)$ starts from $A/p=5/3$ and
has a positive tangent at $T=0$.
This tangent becomes negative for very low $T$ and the anomalous decrease of
$l$ with increasing $T$ takes place; this reminds the situation pictured
in the inset of figure \ref{F1}.
This type of inflection behaviour extends up to $p=1/2$.
As is seen on the anomalous curves for $p=0.5$ and $p=0.4$,
for $1/4<p\le 1/2$ $l(T)$ has a negative tangent even at $T=0$,
so the anomalous NTE behaviour is taking place without an inflection point.
The pressure $p=0.2$ is below $p=1/4$ and therefore has a positive
tangent, not only at $T=0$ but at all $T>0$, and the corresponding $l$
grows monotonously with increasing $T$.

The high-temperature expansion of $l$ is obtained as
\begin{eqnarray} 
l & = & \frac{1}{\beta p} + a + A\left[ (a'-a)+a\ln\left(\frac{a}{a'}\right)
\right] \beta + \frac{A}{2} \Bigg\{ (a'-a)\left[ (3a-a')p-2A \right]
\nonumber \\ & & 
+ 2 a (A-a p)\ln\left(\frac{a'}{a}\right)
+a A \ln^2\left( \frac{a'}{a}\right) \Bigg\} \beta^2 + O(\beta^3) . \label{LnHT}
\end{eqnarray}

To get the virial expansion, one needs a series formula for 
the upper incomplete gamma function \cite{Gradshteyn}:
\begin{equation} \label{Gamma}
\Gamma(x,z) = \Gamma(x) - \frac{z^x}{x} + \frac{z^{1+x}}{1+x}
- \frac{z^{2+x}}{2(2+x)} + O(z^{3+x}) .
\end{equation}
The small-$s$ expansion (\ref{LnOm}) then reads
\begin{eqnarray} 
\widehat{\Omega}(s) & = & \frac{1}{s}-a'+\frac{(a')^2}{2}s
-\frac{(a')^3}{6}s^2 + (a')^{-A\beta}\Bigg[
\frac{(a')^{1+A\beta}-a^{1+A\beta}}{1+A\beta} \nonumber \\ & &
+\frac{a^{2+A\beta}-(a')^{2+A\beta}}{2+A\beta}s
+\frac{(a')^{3+A\beta}-a^{3+A\beta}}{3+A\beta}\frac{s^2}{2}\Bigg]
+ O(s^3) . \label{LnOms}
\end{eqnarray}
The coefficients of the virial expansion (\ref{Virhr}) become
\begin{eqnarray} 
B_2 & = & \frac{a^{1+A\beta} + A\beta\ a'^{1+A\beta}}{(1+A\beta)a'^{A\beta}}
\nonumber\\  
B_3 & = & \frac{1}{(1+A\beta)^2(2+A\beta)^2(a')^{2A\beta}}\nonumber\\
& & \times\biggl\{2(2+A\beta)a^{2+2A\beta}+A\beta[(A\beta)^2+2A\beta-1]
(a')^{2+2A\beta} \nonumber\\
& & - 2[a(1+A\beta)^2-2a'A\beta(2+A\beta)]a^{1+A\beta}(a')^{A\beta}\biggr\} ,
\label{Lnvir}
\end{eqnarray}
etc.

\begin{figure}[t]
\begin{center}
\includegraphics[clip,width=0.84\textwidth]{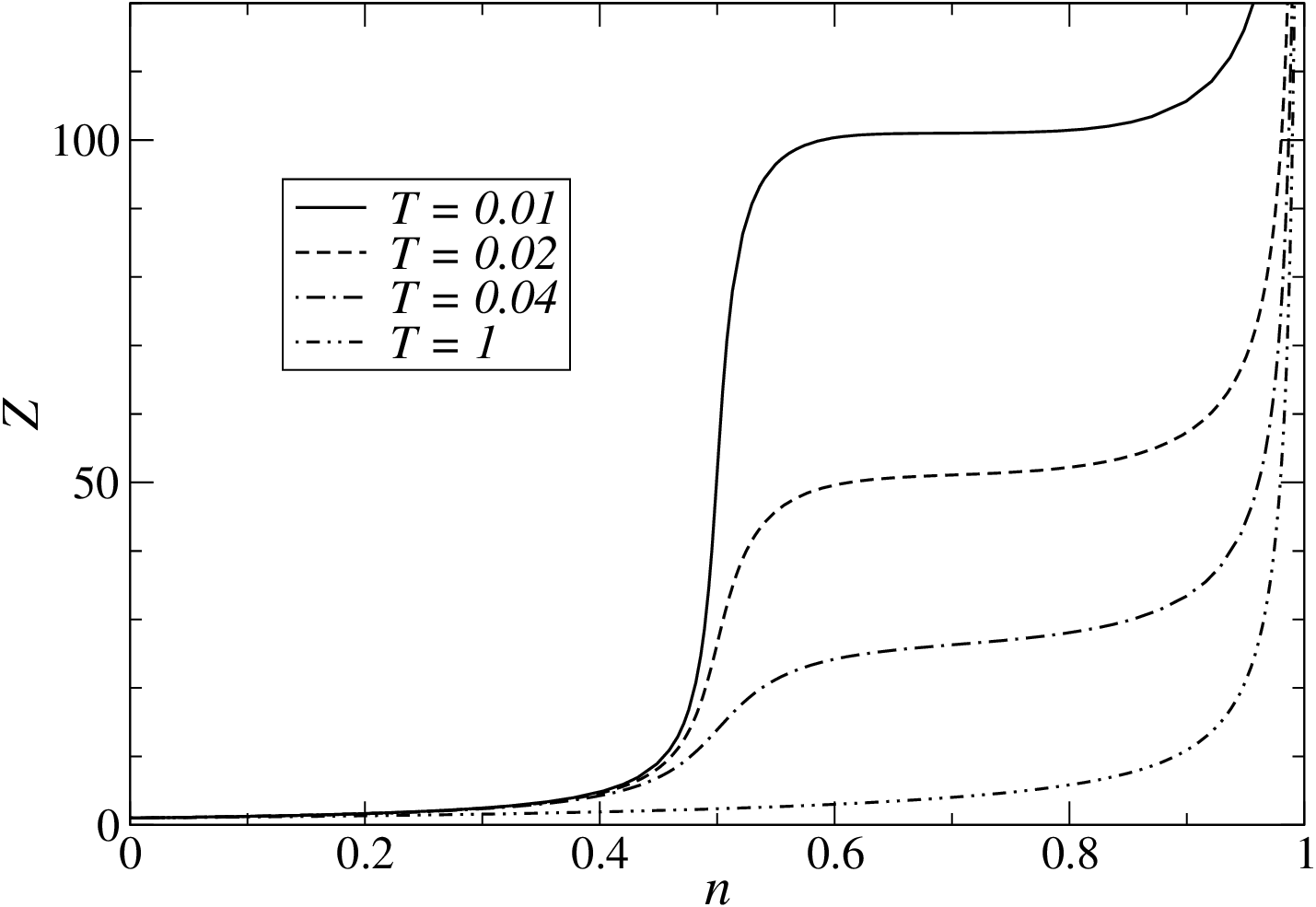}
\caption{The plot of the compressibility factor $Z$ versus
the particle density $n$ for the logarithmic ramp potential (\ref{Ln})
with parameters $a=1$, $a'=2$ and $A=1$.
The isotherms $T=0.01, 0.02, 0.04, 1$ are represented by the solid,
dashed, dash-dotted and dash-double dotted curves, respectively.
See the text for a detailed description of the plots.}
\label{znlog}
\end{center}
\end{figure}

As is documented on isotherms $T=0.01, 0.02, 0.04, 1$
in figure \ref{znlog}, the plot of the compressibility factor $Z$
versus the particle density $n$ {\em always} exhibits the monotonic
growth in the whole interval of $n\in [0,1]$.
This fact confirms that the presence of the NTE anomaly for a given
interaction potential does not automatically imply the nonmonotonic
behaviour of $Z(n)$.

It is instructive to test the asymptotic $a n\to 1^-$ formula for
$Z(n)$ (\ref{Znan}) on the considered logarithmic ramp potential:
\begin{eqnarray}
Z(\beta,n) & = & \frac{1}{1-a n} + \beta A - \beta A (1-a n)
+ O\left[ (1-a n)^2\right] \nonumber \\
& = & \frac{1}{1-n} + \beta - \beta (1-n) + O\left[ (1-n)^2\right] .
\end{eqnarray}
Choosing $n=0.99$ and $\beta=2$, one has successively
\begin{eqnarray}
\frac{1}{Z} \left( Z-\frac{1}{1-n} \right) = 0.0099, \nonumber \\
\frac{1}{Z} \left( Z-\frac{1}{1-n} -\beta \right) = -0.0000492, \nonumber \\
\frac{1}{Z} \left( Z-\frac{1}{1-n} -\beta + \beta(1-n) \right)
= 7.448\times 10^{-7}.
\end{eqnarray}
Thus the relative deviation decreases systematically.

\subsection{Combination of linear and logarithmic ramps} \label{Sec5.2}
Let us consider a model with logarithmic plus linear interaction within
the finite range $a < x < a'$ such that both $\varphi(x)$ and
${\partial\varphi(x)}/{\partial x}$ vanish at $x=a'$:
\begin{equation} \label{Lncd}
\varphi(x)= A \left[ \frac{\vert x\vert}{a'}
-\ln\left( \frac{\vert x\vert}{a'} \right) -1 \right] ,
\qquad a < \vert x\vert < a'.
\end{equation}
For the Laplace transform of interest, one gets
\begin{eqnarray}
\hat\Omega(s) & = & \frac{\e^{-a' s}}{s}
+\e^{A\beta} \left\{
\frac{\Gamma\left[ 1+A\beta,a\left( \frac{A \beta}{a'}+s\right)\right]}{
{a'}^{A\beta} \left( \frac{A \beta}{a'}+s\right)^{1+A\beta}} 
-a' \frac{\Gamma\left[ 1+A\beta,A\beta+a's\right]}{
(A\beta+a's)^{1+A\beta}} \right\} .
\nonumber \\ & & \label{LncdOm}
\end{eqnarray}
To save space, the long formula for $l(\beta,p)$ is not given here.

The low-temperature expansion of $l(T,p)$ reads as
\begin{equation} \label{lLncdf}
l = \left\{ \begin{array}{ll}
\displaystyle{a + \frac{T}{p-A/a'} - \frac{2A T^2}{a^2(p-A/a')^3} + O(T^3)}
& \displaystyle{p>\frac{A}{a}-\frac{A}{a'} ,} \cr
\displaystyle{a + \sqrt{\frac{a'a}{A\pi}} \sqrt{T} + O(T)} &
\displaystyle{p = \frac{A}{a}-\frac{A}{a'} ,} \cr
\displaystyle{\frac{a'A}{a'p+A} + \frac{a'}{a'p+A}T + \cdots} &
\displaystyle{p < \frac{A}{a}-\frac{A}{a'} ,}
\end{array} \right.
\end{equation}
where the dots denote an exponentially small contribution of type
$\exp(-c/T)$.
The incompressibility pressure is given by $p_i=A/a-A/a'$.

In the ground $T=0$ state, the lattice spacing $l(0)=a'A/(a'p+A)$ changes
continuously from $a$ to $a'$ if $p<A/a-A/a'$ and
the formula (\ref{derener}) is applicable in this region of $p$-values.

The prefactors to the leading temperature corrections to the ground state
in (\ref{lLncdf}) are always positive for the considered $a'\le 2a$ which
is a sign of the standard increase of $l$ with $T$, at least in the low-$T$
region.

According to our analysis, $l$ is an increasing function of $T$
for any choice of model's parameters and pressure $p$; the corresponding
figure is omitted for lack of space, the isobaric curves resemble those
in figures \ref{F3} and \ref{F4}.
In other words, the low-$T$ anomaly is absent for
the interaction potential (\ref{Lncd}).

The high-temperature expansion of $l$ takes the form
\begin{eqnarray} 
l & = & \frac{1}{\beta p} + a + \frac{A}{2} \left[ a' - \frac{a^2}{a'}
- 2a \ln\left( \frac{a'}{a}\right) \right] \beta \nonumber \\
& & + \frac{A}{12a'^2} \Biggl\{ (a'-a) \left[ a'a(10a'p+A)+2a^2(2a'p-A)
-a'^2(2a'p+5A) \right] \nonumber \\ & &
+ 6a'a\ln \left( \frac{a'}{a}\right)
\left[ aA-2a'ap + a'A\ln\left( \frac{a'}{a}\right) \right]
\Biggr\} \beta^2 + O(\beta^3) . \label{LncdHT}
\end{eqnarray}

The first nontrivial coefficient of the virial expansion is found to be
\begin{equation} \label{Lncdvir}
B_2 = a' + \frac{a'\e^{A\beta}}{(A\beta)^{1+A\beta}}
\left[ \Gamma(1+A\beta,A\beta) - \Gamma\left( 1+A\beta,\frac{a A\beta}{a'}
\right) \right].
\end{equation}

For any $T$, $Z(n)$ exhibits the monotonic growth in
the whole interval of $n\in [0,1]$.

\subsection{Logarithmic model with fully continuous interaction} \label{Sec5.3}
Next we consider the model with logarithmic interaction within
the range $a < x < a'$ such that $\phi(x\to a^+)\to \infty$
and $\phi(x\to a')\to 0$:
\begin{equation} \label{Lnc}
\varphi(x)= -A \ln \left( \frac{\vert x\vert-a}{a'-a} \right) ,
\qquad a < x < a' .
\end{equation}
The potential is thus the continuous function of the distance in the
whole interval $0<x\le a'$, but its derivative $\varphi'(x)$ is discontinuous
at $x=a'$.

For the Laplace transform (\ref{Om}) one gets
\begin{equation} \label{LnfOm}
\widehat\Omega(s) = \frac{\e^{-a' s}}{s}+\frac{\e^{-a s}}{s}
\frac{\gamma[1+A\beta,(a'-a)s]}{[(a'-a)s]^{A\beta}},
\end{equation}
where we introduced the lower incomplete gamma function
$\gamma(k,z)=\int_0^zt^{k-1}\e^{-t}{\rm d}t$ \cite{Gradshteyn}.
The EoS is found to be
\begin{equation} \label{Lncl}
\hspace{-57pt} l = \frac{\e^{a \beta p}[(a'-a)\beta p]^{A\beta}(1+a\beta p)
+\e^{a' \beta p}(1+A\beta+a\beta p)\gamma[1+A\beta,(a'-a)\beta p]}{
\beta p\{[(a'-a)\beta p]^{A\beta}\e^{a \beta p}+\e^{a' \beta p}
\gamma[1+A\beta,(a'-a)\beta p]\}} .
\end{equation}

\begin{figure}[t]
\centering
\setbox1=\hbox{\includegraphics[clip,height=8.5cm]{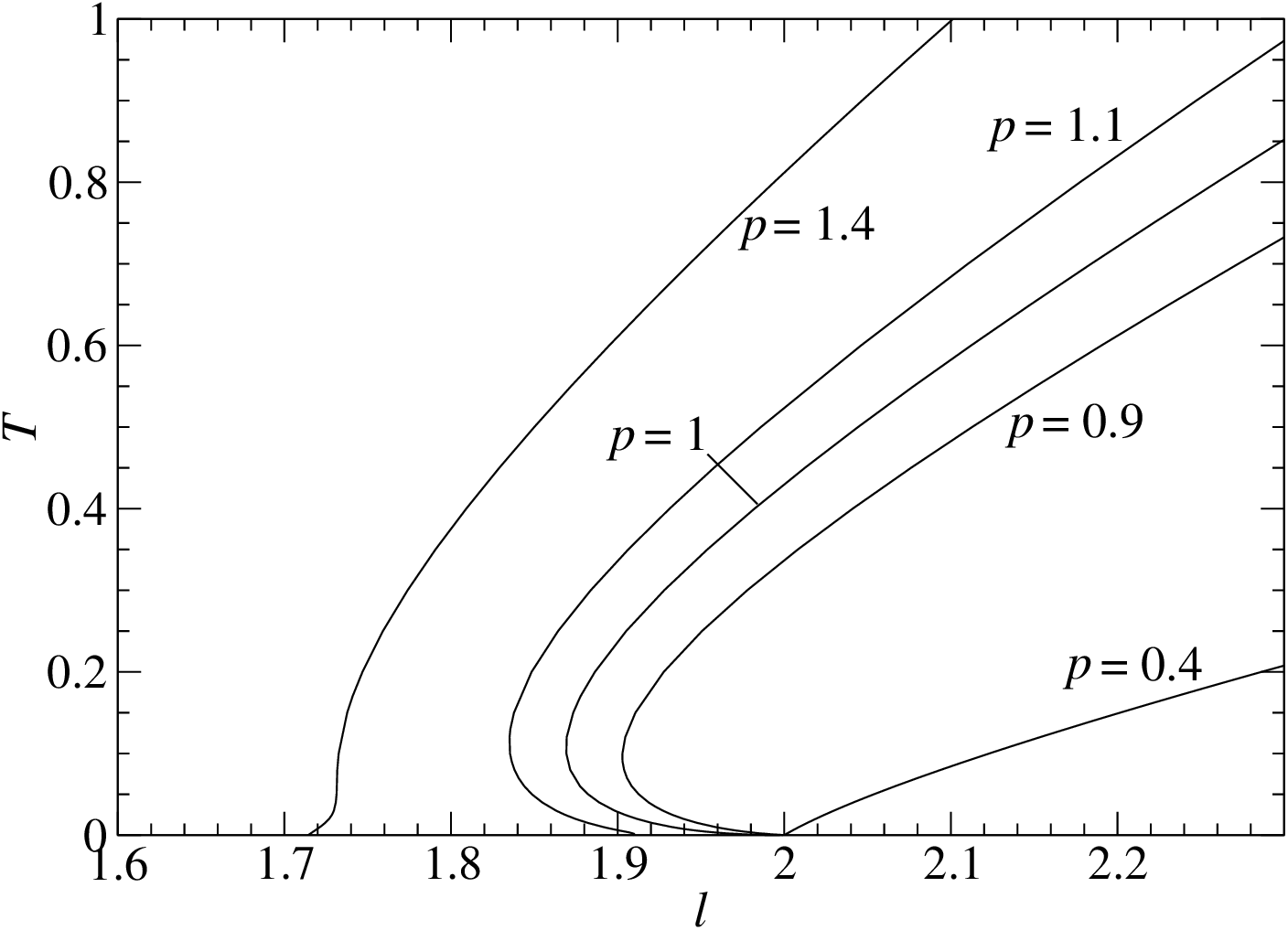}}
\includegraphics[clip,height=8.5cm]{fig9.eps}\lapbox[5cm]{-10.6cm}
{\raisebox{5.1cm}{\includegraphics[clip,height=3cm]{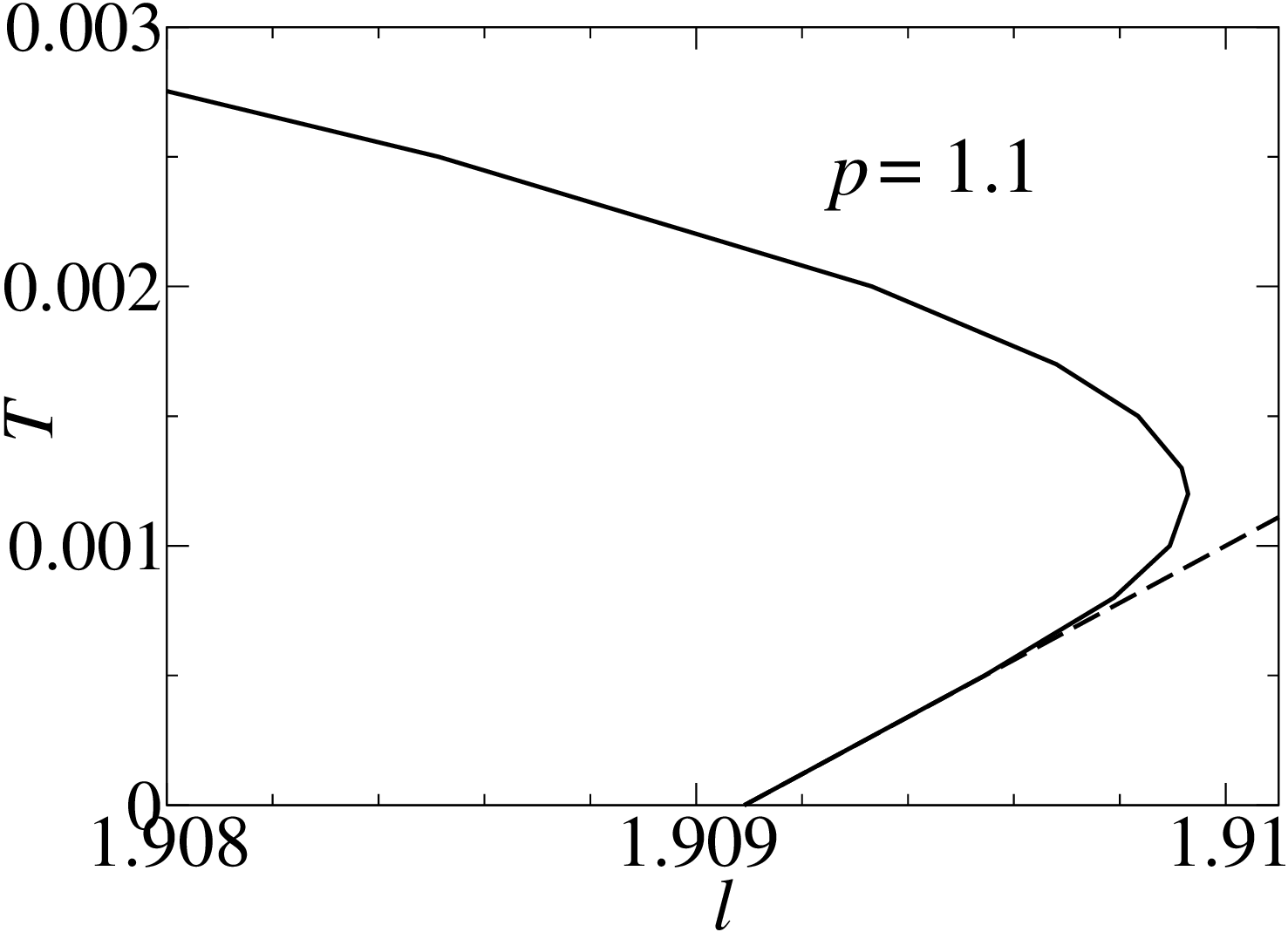}}}
\caption{The plot of $T$ versus $l$ for the continuous logarithmic model
(\ref{Lnc}) with parameters $a=1$, $a'=2$ and $A=1$,
for five values of the pressure $p=1.4,1.1,1,0.9,0.4$.
See the text for a detailed description of the plots.}
\label{F6}
\end{figure}

The low-$T$ expansion of $l(T,p)$ is obtained as
\begin{equation} \label{lLnf}
l = \left\{
\begin{array}{ll}
\displaystyle{a + \frac{A}{p} + \frac{T}{p} + O\left( \e^{-[(a'-a)p-A]/T} \right)}
& \mbox{if $\displaystyle{p>\frac{A}{a'-a}}$,} \cr
\displaystyle{a'-\frac{1}{p}\sqrt{\frac{2A}{\pi}}\sqrt{T} + O(T)} &
\mbox{if $\displaystyle{p=\frac{A}{a'-a}}$,} \cr
\displaystyle{a'+\frac{A-2(a'-a)p}{A-(a'-a)p} \frac{T}{p} + O(T^2)} &
\mbox{if $\displaystyle{p<\frac{A}{a'-a}}$.}
\end{array} \right.
\end{equation}
As a consequence of the divergence $\varphi(x\to a^+)\to\infty$
the incompressibility pressure $p_i\to\infty$.

The ground-state spacing changes itself continuously from $a$ at $p\to\infty$
to $a'$ at $p=A/(a'-a)$; this result can be deduced also from the formula
(\ref{derener}).

The leading corrections to the ground-state values of $l$ are linear in $T$
for $p\ne A/(a'-a)$ and of order $\sqrt{T}$ for $p=A/(a'-a)$.
The prefactor to $T$ is positive for all $p>A/(a'-a)$, the prefactor
to $\sqrt{T}$ is negative for $p=A/(a'-a)$ and finally the prefactor
to $T$ is negative for $A/[2(a'-a)]<p<A/(a'-a)$ and it is positive
for $p<A/[2(a'-a)]$.

Plots of $T$ versus $l$ obtained from the EoS (\ref{Lncl})
with potential parameters $a=1$, $a'=2$ and $A=1$, for five values of
the pressure $p=1.4,1.1,1,0.9,0.4$, are pictured in figure \ref{F6}.
The solution of equations for the inflection point
(\ref{inflection}) reads as
\begin{equation}
p^*(1,2,1) = 1.394097490\ldots , \quad
T^*(1,2,1) = 0.060724189\ldots ;
\end{equation}
$p^*(a,a',A)$ and $T^*(a,a',A)$ can be obtained for arbitrary values of
parameters $a$, $a'$ and $A$ by using (\ref{other}) again.
If $p>p^*$, $l$ grows monotonously with increasing $T$, as is seen
in figure \ref{F6} for $p=1.4$.
In the interval $A/(a'-a)<p<p^*$, or equivalently $1<p<1.394097490\ldots$,
the plot of $T$ versus $l$ is NTE anomalous with a positive tangent in a tiny
interval of low temperatures, see the inset of figure \ref{F6} for $p=1.1$.
The plot of $T$ versus $l$ remains to be NTE anomalous for
$A/[2(a'-a)]<p\le A/(a'-a)$, or $1/2<p<1$, but with a
``normal'' negative tangent at low temperatures,
see the plots for $p=1$ and $p=0.9$ in figure \ref{F6}.
Finally, $l$ grows monotonously with $T$ if $p<A/[2(a'-a)]=1/2$,
as is seen for $p=0.4$.

The high-temperature expansion of $l$ becomes
\begin{equation} \label{LncHT}
l = \frac{1}{\beta p}+a+(a'-a)A \beta - \frac{A}{2} (a'-a) [2A+(a'-a)p]
\beta^2 + O(\beta^3) .
\end{equation}

Using that $\gamma(x,z)=\Gamma(x)-\Gamma(x,z)$ and the series expansion
(\ref{Gamma}), one gets the first virial coefficients:
\begin{eqnarray} \label{Lncvir}
B_2 & = & \frac{a+a'A\beta}{1+A\beta} , \nonumber \\  
B_3 & = & \frac{2a^2+2aa'A\beta(3+A\beta)+(a')^2
A\beta[(A\beta)^2+2A\beta-1]}{(1+A\beta)^2(2+A\beta)} ,
\end{eqnarray}
etc.

For any $T$, the plot of $Z(n)$ exhibits the monotonic
growth in the whole interval of $n\in [0,1]$.
Once more, in close analogy with the logarithmic ramp discussed
in section \ref{Sec5.2}, the presence of the low-temperature NTE anomaly 
does not imply the nonmonotonic behaviour of $Z(n)$.

\begin{figure}[t]
\begin{center}
\includegraphics[clip,width=0.84\textwidth]{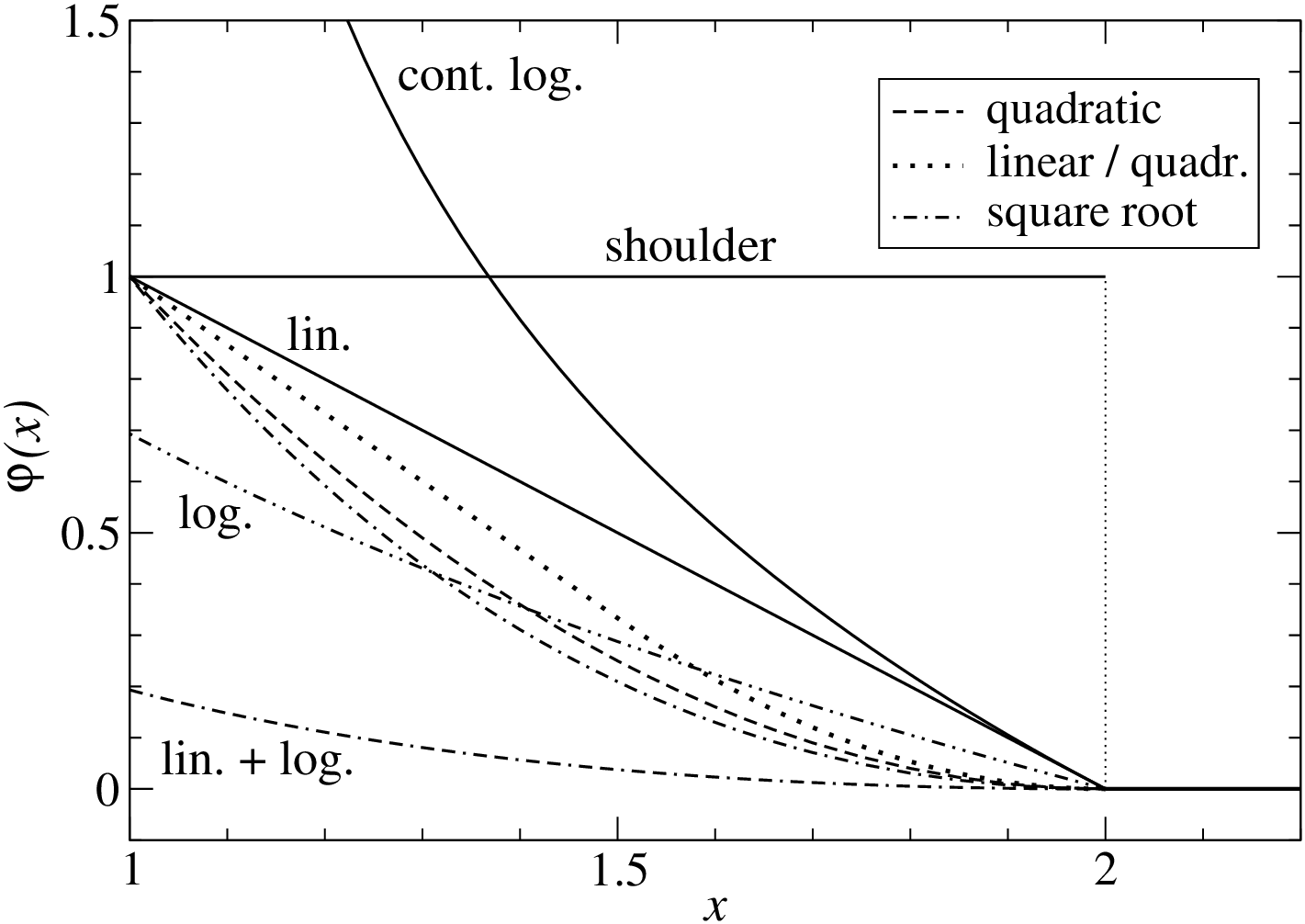}
\caption{The plots of the interaction potentials studied in this paper,
with hard core radius $a=1$ and the finite range $a'=2$.
The square shoulder (denoted as shoulder) and the linear ramp (lin.)
correspond to solid curves, the quadratic ramp (quadratic)
to the dashed curve, the fusion of linear and quadratic ramps
(linear/quadr.) to the dotted curve, the ramp quadratic in
the square root of distance (square root) to the dash-dotted curve,
the logarithmic potential of the ramp type (log.) to the dash-double
dotted curve, the combination of linear and logarithmic ramps
(lin.+log.) to the double dash-dotted curve and the continuous
logarithmic potential (cont. log.) to another solid curve.}
\label{pot}
\end{center}
\end{figure}

\section{Conclusion} \label{Sec6}
Various 1D models of classical particles studied in this work
possess pairwise interaction potentials $\phi(x)$ of repulsive type with
an infinite hard-core component of diameter $a$ and a soft component
of finite range $a'$, see figure \ref{pot}.
Having in mind the NTE, these two-length-scales potentials can be divided
into two groups.
The quadratic ramp (section \ref{Sec4.2}), the ramp quadratic in
the square root of distance (section \ref{Sec4.4}) and
the combination of linear and logarithmic ramps (section \ref{Sec5.2}),
belonging to the first group $A$, do not exhibit NTE.
The square shoulder model (section \ref{Sec3}), the linear ramp
(section \ref{Sec4.1}), the square-root ramp 
(section \ref{Sec4.4}), the logarithmic ramp (section \ref{Sec5.1}) and
the continuous logarithmic potential (section \ref{Sec5.3}),
belonging to the second group $B$, do exhibit NTE.

\begin{figure}[t]
\begin{center}
\includegraphics[clip,width=0.84\textwidth]{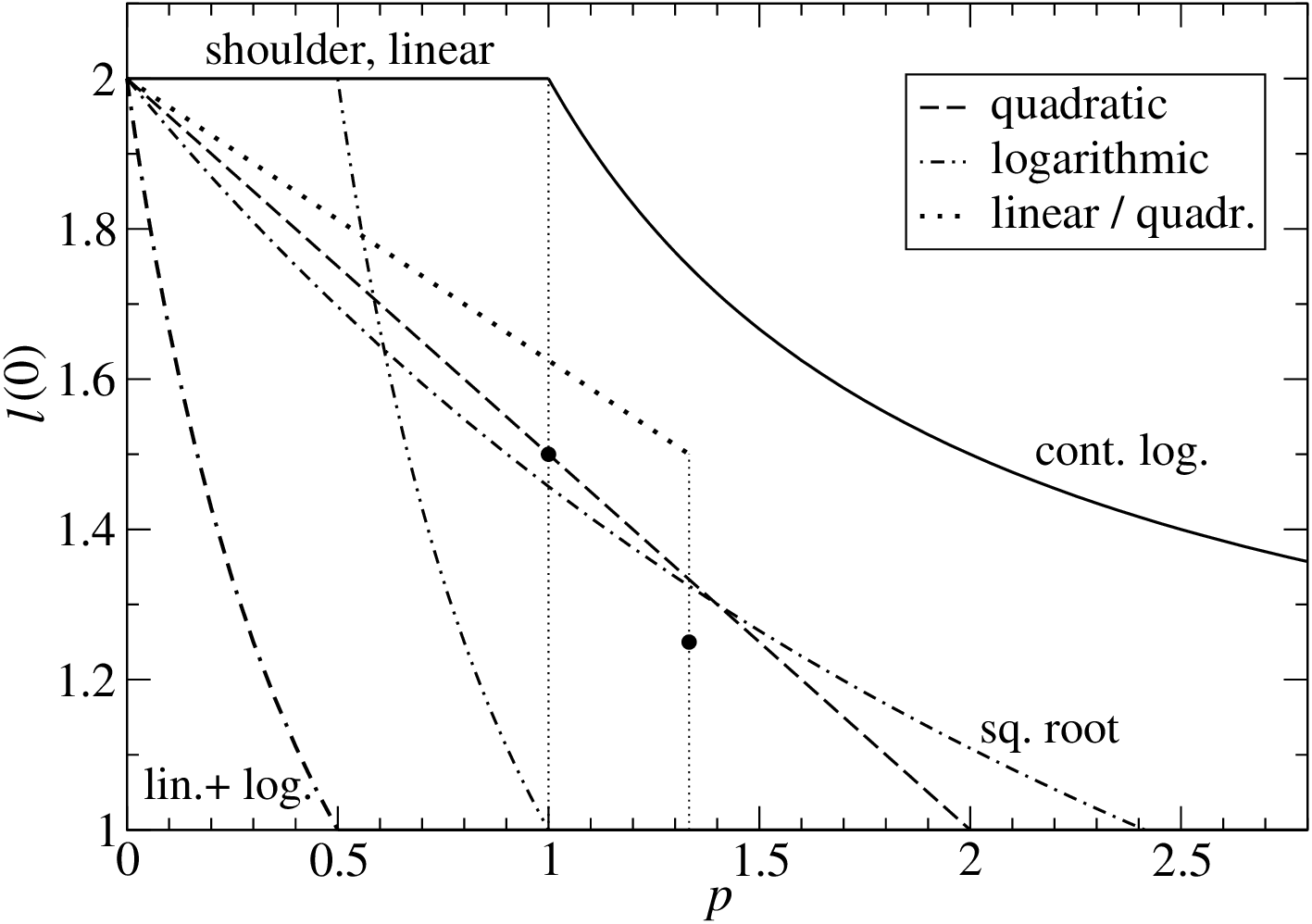}
\caption{The spacing of the array of particles in the ground state
$l(0)\equiv l(T=0,p)$ versus the applied pressure $p$ for models studied
in this paper, with parameters of the interaction potential fixed to $a=1$,
$a'=2$ and $\varepsilon=1$ or $A=1$.
The square shoulder (section \ref{Sec3}) and the linear ramp
(section \ref{Sec4.1}) correspond to the horizontal solid lines
$l(0)=2$ for $p<1$ and $l(0)=1$ for $p>1$, interconnected discontinuously
(vertical dotted line) through the point $l(0)=3/2$ at $p=1$.
The quadratic ramp (section \ref{Sec4.2}) is described by the dashed
curve for $p<2$ and the horizontal $l(0)=1$ solid line for $p>2$.
The fusion of linear and quadratic ramps (section \ref{Sec4.3}) is
represented by dotted curve followed by a jump of the spacing to 1 at $p=4/3$.
The ramp quadratic in the square root of distance (section \ref{Sec4.4})
corresponds to the dash-dotted curve for $p<1/(\sqrt{2}-1)=2.41421\ldots$
and to the horizontal $l(0)=1$ solid one for $p>2.41421\ldots$.
The logarithmic potential of the ramp type (section \ref{Sec5.1})
corresponds to the horizontal solid line $l(0)=2$ for $p<1/2$, to the
dash-double dotted curve for $1/2<p<1$, the last is intertwined
with the horizontal solid line $l(0)=1$ for $p>1$.
The combination of linear and logarithmic ramps (section \ref{Sec5.2})
is represented by the double dash-dotted curve for $0<p<1/2$ and
by the horizontal solid line $l(0)=1$ for $p>1/2$.
The model with the continuous logarithmic potential (section \ref{Sec5.3})
is represented by the horizontal solid line $l(0)=2$ for $p<1$,
linked with another solid curve $1+1/p$ for $p>1$.}
\label{F7}
\end{center}
\end{figure}

The quantity of primary interest is the reciprocal density $l(T,p)$.
As soon as $a'\le 2a$, the pair interaction reduces itself effectively
to nearest neighbours which permits a closed-form solution (\ref{recdensity})
for $l$ in terms of the Laplace transform of the pair Boltzmann factor. 
All models studied in this paper admit an explicit expression for
$l(T,p)$ in terms of elementary or special functions.
For a fixed $p$, $T\to 0$ $(\beta\to\infty)$ limit of this formula
yields the ground equidistant state of particles and the leading
low-temperature correction to the ground state.

The equidistant arrays of particles in the ground state, taken with
parameters of the interaction potential fixed to $a=1$, $a'=2$ and
$\varepsilon=1$ or $A=1$, are summarized in figure \ref{F7}.
The three potentials of group $B$, namely the square shoulder,
the linear ramp and the fusion of linear/quadratic ramps,
have discontinuity in ground states which skip from spacing
$l(0)=a=1$ for $p>p_i$ ($p_i$ denotes the incompressibility pressure)
to $l(0)=a'=2$ for $p<p_i$, via an intermediate one
with the mean spacing $l(0)=(a+a')/2=3/2$ at $p=p_i$.
The formula (\ref{derener}) for zero temperature lattice constant $l(0)$
is applicable in all other cases when $l(0)$ is within the interval
$a<l(0)<a'$.
The equidistant ground states of the quadratic ramp and logarithmic
models exhibit a continuous change of chain spacing from $a$ to $a'$
as $p$ changes from $\infty$ to $0$.

The incompressibility pressure $p_i=\varepsilon/(a'-a)$ for the square shoulder 
potential (\ref{SW}), $p_i=\varepsilon/(a'-a)$ for the linear ramp
(\ref{lrm}), $p_i=2 \varepsilon/(a'-a)$ for the quadratic ramp
(\ref{Q}), $p_i=4/3$ for fusion of linear/quadratic ramps
(\ref{lqphi}), $p_i=\varepsilon/(\sqrt{a' a}-a)$ for the square-root ramp
(\ref{QSR2}), $p_i=A/a$ for the logarithmic ramp (\ref{Ln}),
$p_i=A/a-A/a'$ for the combination of linear and logarithmic ramps
(\ref{Lncd}) and $p_i\to\infty$ for the continuous
logarithmic interaction (\ref{Lnc}).
The formula (\ref{derener}) tells us that this pressure is given by
\begin{equation}
p_i = - \frac{\partial\varphi(x)}{\partial x}\Bigg\vert_{x=a} .
\end{equation}
It can be checked that every interaction potential fulfills this relation,
except for the square shoulder (probably because of the discontinuity
of the potential at $x=a'$).

The analytical structure of the low-$T$ expansion of $l$ is modified
by passing through special values of the pressure $p$.
Low-temperature series splits into 3-5 distinct cases according to
the relation between the pressure and model's parameters. 
An important information taken from the leading $T$-correction to the
ground state is its positive or negative sign.
The positive sign indicates the expected increase of $l$ with $T$
while the negative sign signalizes the anomalous NTE decrease of $l$ with $T$,
at least in the low-$T$ region.
The sign of the leading $T$-correction is always positive for models of
group $A$, see equation (\ref{lQ}) for the quadratic ramp, equation
(\ref{lQSRLT}) for the square-root ramp and equation (\ref{lLncdf})
for the combination of the linear and logarithmic ramps.
Exact results for models of group A, pictured in figures \ref{F3}
and \ref{F4}, reveal that the anticipated increase of $l$ with $T$ takes
place for any value of $p$, i.e., they are free of the NTE anomaly.
On the other hand, models of group $B$ exhibit in certain intervals of
$p$-values the NTE anomaly (at low enough temperatures).
As is documented by insets in figure \ref{F1} for the square shoulder model
and in figure \ref{F6} for the continuous logarithmic model, the positive sign
of the leading $T$-correction of $l$ for models of group $B$ does not
exclude an anomalous NTE plot of $l(T)$ for higher (but still relatively
low) strictly positive temperatures.

Here is a brief summary of the relevant output of the present paper
in view of the previous approaches.

The previous evidence for the low-temperature NTE anomaly
in 2D and 3D  systems was related to the anisotropy of the intermolecular
potential, special topography of the energy landscape for isotropic
potentials (with ground states ranging from fully disordered to
crystalline ones) or two-length-scales nature of isotropic repulsive
potentials. 
The results were obtained based on experimental measurements and numerical
analysis of data by using various simulation methods.
Here, the NTE phenomenon is studied within 1D fluids with nearest neighbour
interactions whose thermal equilibrium is exactly solvable.
The exact treatment of various types of two-length-scales potentials
shows that the presence of the NTE phenomenon depends very much on
the shape of the interaction potential; this puts into doubts
the previous hypothesis that the two-length-scales repulsive
potentials always exhibit water-like anomalies \cite{Oliveira09},
at least in 1D.

A precise mathematical specifications of potential's
characteristics which lead to the NTE anomaly remains to be an open problem.
It was shown by Gribova et al. \cite{Gribova09} that the NTE phenomenon
exists also for a fully analytic potential similar to the repulsive square
shoulder one.
The necessary and sufficient condition for NTE of Kuzkin (\ref{Kuzkin})
does not apply in general to the present 1D potentials.
For instance, for the logarithmic ramp potential (\ref{Ln}), which
exhibits the NTE anomaly, the third derivative $\varphi'''(x)=-2A/x^3$
is negative.
This might be caused by the nonanalytic structure of the considered
interaction potentials; the third derivative is not well defined at points
where the potential or one of its first two derivatives is discontinuous.

As concerns the present case of 1D fluids, the lattice
spacing of equidistant ground states can skip discontinuously at the
incompressibility pressure $p_i$ as the pressure varies.
The analytical structure of the low-$T$ expansion of the inverse density
$l(T,p)$ depends on ranges of $p$-values.

We have also studied whether the NTE phenomenon has
some impact on the plot of the compressibility factor $Z$ (\ref{Z}) as
the function of the particle density $n$, at fixed temperature $T$.
For potentials of the first group A, which are free of the NTE phenomenon,
the plot $Z(n)$ always exhibits the monotonic growth in the whole
interval of $n\in [0,1]$.
For potentials of the second group B, which exhibit the NTE anomaly,
the plot of $Z$ versus $n$ can exhibit both the nonmonotonic behaviour
with one maximum/minimum at low enough temperatures, see figure \ref{znsh}
for the square shoulder potential, as well as the monotonic growth,
see figure \ref{znlog} for the logarithmic ramp potential (\ref{Ln}).
This means that the relationship between the NTE anomaly and the nonmonotonic
behaviour of the compressibility factor $Z$ was not confirmed.

Another open problem for future is an extension of the 1D repulsive analysis
to purely attractive or mixed repulsive-attractive interaction potentials.
The present study can be directly extended to mixtures, say
the polydisperse Tonks gas \cite{Santos16,Evans10}.

Another possibility is the exact treatment of 1D quantum fluids.
Degenerate Bose version of these models can be realized experimentally
via a strong confinement, e.g., in optical lattices
\cite{Kinoshita04,Haller10} and micro-traps \cite{Kruger10},
see chapter 2 {\it Experimental realization of one-dimensional Bose gases}
of monograph \cite{Langen15}.
The crucial theoretical fact is that even quantum particles cannot tunnel
over an infinite potential barrier of the hard core.
The coordinate and thermodynamic Bethe-ansatz solution of Tonks gas
\cite{Wadati02,Samaj13} can serve as a starting point for an exact analysis
of a 1D quantum fluid with nearest-neighbour interactions; the best candidate
is the square shoulder model.
  
\ack
The support received from the project EXSES APVV-20-0150 and VEGA Grant
No. 2/0089/24 is acknowledged.

\end{document}